\documentclass[]{pasj01}
\draft

\Received{}%{yyyy/mm/dd}
\Accepted{}%{yyyy/mm/dd}
\newcommand{\actaa}{AcA }
\newcommand{\na}{NewA.  }

\begin{document}

\title{Comprehensive Photometric Investigation of an Active Early K-type Contact System -- IL Cnc
}

%%% begin:list of authors
% Do NOT capitalize all letters in "textsc".
\author{N.-P. \textsc{Liu}\altaffilmark{1,3},
        T. \textsc{Sarotsakulchai}\altaffilmark{2,4},
        S. \textsc{Rattanasoon}\altaffilmark{2,5},
        B. \textsc{Zhang}\altaffilmark{6,7}}

\altaffiltext{1}{Yunnan Observatories,
Chinese Academy of Sciences, P.O. Box 110, Kunming 650216,
P.R.China}
\email{lnp@ynao.ac.cn}
\altaffiltext{2}{National Astronomical Research Institute of Thailand,
260 Moo 4, T.Donkaew, A.Maerim, Chiangmai, 50180, Thailand}
\altaffiltext{3}{Key Laboratory for
the Structure and Evolution of Celestial Objects, Chinese Academy of
Sciences, P.O. Box 110, Kunming 650216, P.R.China}
\altaffiltext{4}{University of
Chinese Academy of Sciences, Yuquan Road 19$^{\#}$, Shijingshan Block, Beijing 100049, P.R.China}
\altaffiltext{5}{Department of Physics and Astronomy, The University of Shelffield, Hounsfield Road, Sheffield S3 7RH, UK}
\altaffiltext{6}{Guizhou Normal University, Guiyang 550001, P.R.China}
\altaffiltext{7}{Guizhou Provincial Key Laboratory of Radio Astronomy and Data Processing, Guiyang 550001, P.R.China}
%%% end:list of authors

%% `\KeyWords{}' always has to be placed before ``\maketitle'' 
%%  List of Key Words:  https://academic.oup.com/pasj/pages/Pasj_Keywords 
\KeyWords{binaries : close --
          binaries : eclipsing --
          stars: evolution --
          stars: individual (IL Cnc)}

\maketitle

\begin{abstract}
Comprehensive photometric investigation was carried out to the early K-type contact binary - IL Cnc. A few light curves from both ground-based telescopes and the Kepler space telescope were obtained (or downloaded) and then analyzed in detail. They are mostly found to be asymmetric and there are even continuously changing O'Connell effect in the light curves from Kepler K2 data, suggesting the system to be highly active. Using the Wilson-Devinney code (version 2013), photometric solutions were derived and then compared. It is found that the calculation of the mass ratio is easily affected by the spot settings. Combining the radial velocities determined from LAMOST median resolution spectral data, the mass ratio of the binary components is found to be $M_2/M_1 = 1.76\pm 0.05$. The components are in shallow contact ($f\sim 9\%$) and have a temperature difference about $T_2 - T_1 = -280\pm 20$ K. The system is demonstrated to be W-subtype, which may be a common feature of K-type contact binaries. The masses of the binary components were estimated to be $M_1\sim 0.51$ M$_{\solar}$ and $M_2\sim 0.90$ M$_{\solar}$. The values are in good agreement with that deduced from the parallax data of Gaia. The results suggest that the primary component lacks luminosity compared with the zero main sequence. The H$\alpha$ spectral line of the primary component is found to be peculiar. Combining newly determined minimum light times with those collected from literature, the orbital period of IL Cnc is studied. It is found that the (O$-$C)s of the primary minima show sinusoidal variation while the secondary do not. The oscillation is more likely to be caused by the starspot activities. Yet this assumption needs more data to support. 
\end{abstract}

\section{Introduction}

K-type contact binaries are important objects because of their special properties \citep{Bradt85} such as O'Connell effect (unequal heights of light maximum, see \cite{Oconn51a}, \yearcite{Oconn51b}), period cut-off \citep{Paczynski06,Rucinski08,LiK19}, W-subtype phenomenon (referring to W- and A-type classifications of contact binaries, see \citet{Binnk70}) and etc.. A recent statistical study shows that the period distribution of EW type eclipsing binaries peaks at 0.29 days \citep{Qian17} which is quite close to the period range of early K-type contact systems. Combining their special properties, the early K-type contact systems are key to the study of short period contact binaries. With fast rotation and temperature similar to the sun, the early K-type contact systems are supposed to be active. IL Cnc is such an early K-type system which might have these special properties that come into sight.

IL Cnc (cross id: GSC 1400-0455, EPIC 211988016, and etc.. $\alpha_{2000.0}$ = $08^{h}55^{m}51^{s}.49$, $\delta_{2000.0}$ = $+20^{\circ}03'38''.6$) was first discovered as an EW-type variable by \citet{Rinner03} based on unfiltered ccd data \citep{Alton18}. Photometric data were also collected by surveys such as ROSTE-I survey (NSVS, \cite{Wozniak04}) and ASAS survey \citep{Pojmi05}. With a short period about only 0.267 days and a large amplitude of 0.6 mag, this object was monitored a number of times for minimum light times. Just recently, this target was studied by \citet{Alton18} through analyzing light curves from the 0.28-m  telescope in UnderOak Observatory and the 0.4-m telescope in Desert Bloom Observatory. Using PHOEBE 0.31a, the author found a preliminary result of roughly determined mass ratio about 1.5 to 2.0, but not yet conclusive determined photometric elements, nor any ``meaningful'' changes in its orbital period. The author also found this target an active system with asymmetric maxima in its light curves. More information about this target has been published thanks to large surveys including Kepler mission \citep{Boruki10}, LAMOST (Large Sky Area Multi-Object Fiber Spectroscopic Telescope, e.g. \cite{CuiX12,Zhao12}) and Gaia \citep{gaia16}. Therefore, it is possible to study this target in more detail.

\section{Photometric Observations from ground-based telescopes}\label{secobs}

\begin{figure*}
\begin{center}
\includegraphics[angle=0,scale=0.85]{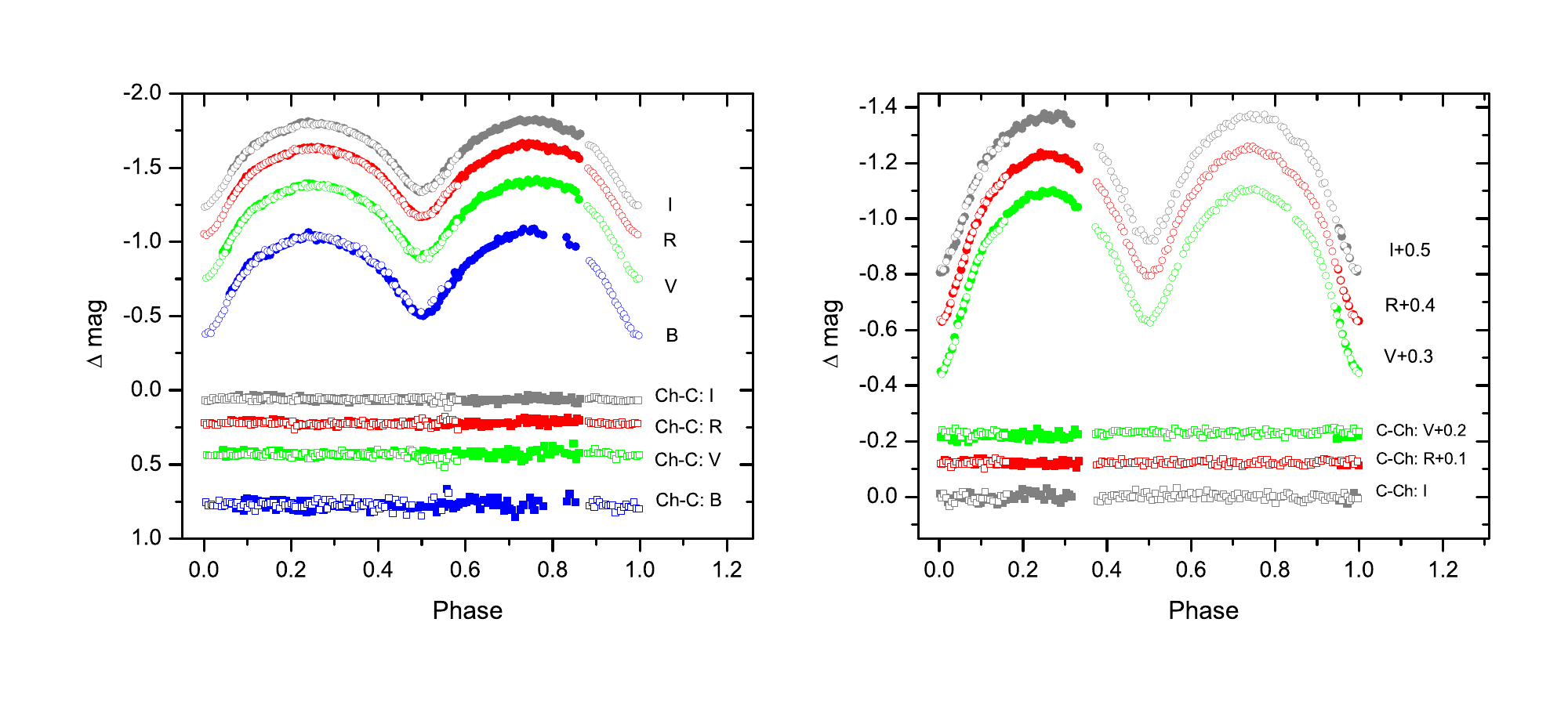}
\caption{Light curves of IL Cnc obtained from TNO 2.4-m telescope (left panel) and 70 cm telescope in Lijiang (right panel).
         The (C-Ch)s or (Ch-C)s (light curve differences between the comparison and the check stars) are plotted at the bottom.
         The filled and open symbols denote data obtained on the first and second night of each set and the colors of 
         the symbols represent the corresponding filters (except grey for the I filter)}\label{figlcs}
\end{center}
\end{figure*}

Complete CCD photometric observations of IL Cnc were carried out first on April 04 and 05, 2015, using the 2.4-m telescope of TNO (Thai National Observatory), which is located in Chiangmai, Thailand, and operated by NARIT (National Astronomical Research Institute of Thailand). Four filters B,V,R and I in Johnson Cousin system were used together with the ARC 4K CCD which has a field of view $8.8' \times 8.8'$ and the integration time was 10 s for each image. In order to make a comparative analysis, the target was subsequently observed on November 16 and 17, 2016, using the sino-Thai 70 cm telescope in Lijiang, China. This time only three filters (VRI) were put to use with a larger field of view about $20.5' \times 20.5'$, since the Andor technology 2K CCD was used. The exposure times were changed to 60 s (V), 40 s (R) and 30 s (I). Both sets of data were reduced in a standard way using ccdred packages from IRAF. A differential photometry was applied in order to get the net variation (light curves) of the target. The standard deviations of the C-Ch data were calculated to be 0.021 mag (B), 0.016 mag (V), 0.013 mag (R) and 0.012 mag (I) for the first set, and 0.009 mag (V), 0.007 mag (R),and 0.012 mag (I) for the second set. The original data of these two sets of light curves can be retrieved from online material. The two sets of light curves were phased using the ephemerides
\begin{displaymath}
   \mathrm{min.I~(HJD)} = 2457118.08544(8)+0^{\mathrm{d}}.267656\times~ E
\end{displaymath}
   and
\begin{displaymath}
   \mathrm{min.I~(HJD)} = 2457710.40859(12)+0^{\mathrm{d}}.267656\times~ E,
\end{displaymath}
respectively. The phased light curves are shown in Figure \ref{figlcs}. The O'Connell effect is clearly seen (about $0.02\sim0.04$ mag.) in the first set of light curves. Some basic information of the variable, comparison and check stars is listed in Table \ref{tabcoo}.

\begin{table}[htbp]
\caption{Basic information of the variable (IL Cnc), the comparison (C) and check (Ch) stars from Vizier database\textsuperscript{\ref{ftvizier}}.}\label{tabcoo}
\begin{center}
\begin{tabular}{lll}
\hline
Target  &  Vmag  &  J-H \\\hline
IL Cnc (V) & 12.63 & 0.452 \\
2MASS J08554788+2001564 (C)  & 13.95 & 0.524 \\
2MASS J08554174+2003285 (Ch) & 14.36 & 0.228 \\
\hline
\end{tabular}
\end{center}
\end{table}

Aside from those light curves, a few minimum light times were determined by using a least-squares parabolic fitting method. A few of them were determined by using the data from 60 cm telescope of YNOs in Kunming, China. The newly determined minimum light times are listed in Table \ref{tabocn}.

\begin{table}[htbp]
\caption{Newly determined minimum light times of IL Cnc.}\label{tabocn}
\footnotesize
\begin{center}
\begin{tabular}{@{}lcccrc}
\hline
\multicolumn{1}{c}{HJD} &\multicolumn{1}{c}{Err}  & Min & Band & NA  &  Source  \\
    2,400,000+          &\multicolumn{1}{c}{(days)} &    &     &  & \\
\hline
57117.14918 &  0.00020  &   II  &  B  &  26 &  CM2.4m  \\
57117.14875 &  0.00015  &   II  &  V  &  21 &  CM2.4m  \\
57117.14884 &  0.00015  &   II  &  R  &  25 &  CM2.4m  \\
57117.14868 &  0.00014  &   II  &  I  &  23 &  CM2.4m  \\
57118.08540 &  0.00016  &   I   &  B  &  22 &  CM2.4m  \\
57118.08535 &  0.00017  &   I   &  V  &  22 &  CM2.4m  \\
57118.08539 &  0.00013  &   I   &  R  &  22 &  CM2.4m  \\
57118.08541 &  0.00015  &   I   &  I  &  22 &  CM2.4m  \\
57118.21799 &  0.00056  &   II  &  B  &  29 &  CM2.4m  \\
57118.21966 &  0.00023  &   II  &  V  &  28 &  CM2.4m  \\
57118.21884 &  0.00017  &   II  &  R  &  23 &  CM2.4m  \\
57118.21963 &  0.00020  &   II  &  I  &  26 &  CM2.4m  \\
57709.33840 &  0.00022  &   I   &  V  &  19 &  LJ70cm  \\
57709.33841 &  0.00024  &   I   &  R  &  19 &  LJ70cm  \\
57709.33846 &  0.00025  &   I   &  I  &  18 &  LJ70cm  \\
57710.27439 &  0.00028  &   II  &  I  &  24 &  LJ70cm  \\
57710.27444 &  0.00019  &   II  &  R  &  25 &  LJ70cm  \\
57710.27473 &  0.00028  &   II  &  V  &  27 &  LJ70cm  \\
57710.40836 &  0.00022  &   I   &  I  &  25 &  LJ70cm  \\
57710.40869 &  0.00019  &   I   &  R  &  23 &  LJ70cm  \\
57710.40851 &  0.00019  &   I   &  V  &  23 &  LJ70cm  \\
58249.06590 &  0.00020  &   II  &  R  &  40 &  YNO60cm \\
58249.06571 &  0.00020  &   II  &  V  &  42 &  YNO60cm \\
58437.36130 &  0.00020  &   I   &  R  &  29 &  YNO60cm \\
58437.36120 &  0.00029  &   I   &  V  &  30 &  YNO60cm \\
\hline
\end{tabular}
\end{center}
\begin{tabnote}
\footnotemark[] Notes. CM2.4m = 2.4-m telescope in Chiangmai. LJ70cm = 70 cm telescope in Lijiang. YNO60cm = 60 cm telescope of Yunnan observatories. NA is the total number of data used to determine the times of minimum light.
\end{tabnote}
\end{table}

\section{Photometric investigation based on ground-based data}
\subsection{Orbital Period Investigation}\label{secper}

This target has been monitored by several investigators over a time span about 20 years. A few minimum light times were published. We collected all the available times of minimum light from literature and database. They are listed in Table \ref{tabocg}, including four reprocessed data. These four data were recalculated from ROTSE and ASAS data by using an average method (see \cite{Liun15}) because they were observed with long cadence and thus sparsely sampled. The epoch and O$-$C values of all the times of minimum light were calculated using the ephemeris given by O$-$C gateway\footnote{var2.astro.cz/ocgate/\label{ocgw}} as follows:

\begin{equation}\label{eq:ephe}
\mathrm{min.I~(HJD)} = 2452721.5705 + 0^{\mathrm{d}}.267656\times~E
\end{equation}

\begin{table*}
\caption{Collection list of minimum light times of IL Cnc.}\label{tabocg}
\footnotesize
\begin{center}
\begin{tabular}{llrrcccll}
\hline %%\tableline
\multicolumn{1}{c}{HJD} &\multicolumn{1}{c}{Err} & Epoch & \multicolumn{1}{c}{(O$-$C)} &  Min & Method &  Obs. & Ref. & Notes \\
           2,400,000+     &\multicolumn{1}{c}{(days)} &    & \multicolumn{1}{c}{(days)} &   &     &   &   & \\
\hline %%\tableline
51528.49322 &  0.0012 &  -4457.5  &  -0.00066  &  s &  ccd  &  PA  & (1)   & reproc. \\
51578.67834 &  0.0010 &  -4270.0  &  -0.00104  &  p &  ccd  &  PA  & (1)   & reproc. \\
52721.5705  &  0.0008 &      0.0  &   0.00000  &  p &  R    &  WN  & (3)   &         \\
53004.88338 &  0.0015 &   1058.5  &  -0.00100  &  s &  ccd  &  PA  & (2)   & reproc. \\
53065.23884 &  0.0014 &   1284.0  &  -0.00196  &  p &  ccd  &  PA  & (2)   & reproc. \\
54500.4124  &  0.0004 &   6646.0  &   0.00012  &  p &  -Ir  &  RM  & (4)   &         \\
54831.9068  &  0.0009 &   7884.5  &   0.00257  &  s &  V    &  DR  & (5)   & unused  \\
54866.4299  &  0.0003 &   8013.5  &  -0.00196  &  s &  -U-I &  RM  & (6)   &         \\
55245.8286  &  0.0009 &   9431.0  &  -0.00564  &  p &  V    &  DR  & (7)   &         \\
55275.4110  &  0.0013 &   9541.5  &   0.00078  &  s &  -Ir  &  AF  & (6)   &         \\
55295.3479  &  0.0010 &   9616.0  &  -0.00270  &  p &  -Ir  &  AF  & (6)   &         \\
55295.4840  &  0.0009 &   9616.5  &  -0.00042  &  s &  -Ir  &  AF  & (6)   &         \\
55523.9260  &  0.0002 &  10470.0  &  -0.00282  &  p &  R    &  NR  & (8)   &         \\
55571.8365  &  0.0003 &  10649.0  &  -0.00274  &  p &  V    &  DR  & (9)   &         \\
55571.9700  &  0.0003 &  10649.5  &  -0.00307  &  s &  V    &  DR  & (9)   &         \\
55627.3762  &  0.0002 &  10856.5  &  -0.00166  &  s &  -U-I &  RM  & (10)  &         \\
55667.6576  &  0.0004 &  11007.0  &  -0.00249  &  p &  V    &  DR  & (9)   &         \\
56000.6190  &  0.0040 &  12251.0  &  -0.00516  &  p &  V    &  DR  & (11)  &         \\
56000.7575  &  0.0007 &  12251.5  &  -0.00048  &  s &  V    &  DR  & (11)  &         \\
56355.6678  &  0.0002 &  13577.5  &  -0.00204  &  s &  ccd  &  NR  & (12)  &         \\
56643.5313  &  0.0001 &  14653.0  &  -0.00257  &  p &  ccd  &  MW  & (13)  &         \\
56677.7910  &  0.0002 &  14781.0  &  -0.00284  &  p &  ccd  &  NR  & (14)  &         \\
56711.6489  &  0.0003 &  14907.5  &  -0.00342  &  s &  BVIc &  AK  & (15)  &         \\
56714.5936  &  0.0003 &  14918.5  &  -0.00294  &  s &  BVIc &  AK  & (15)  &         \\
56719.1427  &  0.0002 &  14935.5  &  -0.00399  &  s &  BVIc &  AK  & (15)  &         \\
56720.6151  &  0.0006 &  14941.0  &  -0.00370  &  p &  BVIc &  AK  & (15)  &         \\
56732.5252  &  0.0005 &  14985.5  &  -0.00429  &  s &  BVIc &  AK  & (15)  &         \\
56743.3679  &  0.0011 &  15026.0  &  -0.00166  &  p &  -I   &  AF  & (16)  &         \\
56743.5003  &  0.0011 &  15026.5  &  -0.00308  &  s &  -I   &  AF  & (16)  &         \\
57414.3818  &  0.0005 &  17533.0  &  -0.00135  &  p &  -I   &  AF  & (17)  &         \\
57414.5167  &  0.0007 &  17533.5  &  -0.00028  &  s &  -I   &  AF  & (17)  &         \\
58129.8257  &  0.0002 &  20206.0  &  -0.00194  &  p &  BVIc &  AK  & (15)  &         \\
58130.8961  &  0.0002 &  20210.0  &  -0.00216  &  p &  BVIc &  AK  & (15)  &         \\
58131.8318  &  0.0001 &  20213.5  &  -0.00326  &  s &  BVIc &  AK  & (15)  & unused  \\
58131.9667  &  0.0002 &  20214.0  &  -0.00218  &  p &  BVIc &  AK  & (15)  &         \\
58139.1932  &  0.0010 &  20241.0  &  -0.00240  &  p &  V    &  IH  & (18)  & \\
58139.3272  &  0.0010 &  20241.5  &  -0.00222  &  s &  V    &  IH  & (18)  & \\
\hline %%\tableline
\end{tabular}
\end{center}
\begin{tabnote}
\footnotemark[]Notes. Obs.=Observer; AF=Agerer Franz; AK=ALTON, K.B.; IH=Itoh Hiroshi;
DR=Diethelm Roger; MW=Moschner Wolfg; NR=Nelson Robert; PA=Paschke Anton;
 RM=Raetz Manfred; WN=Waelchli Nicolas. The data were collected with the help of O$-$C gateway\textsuperscript{\ref{ocgw}}.\\
Reference:
(1) ROTSE \citep{Gettel06}; (2) ASAS \citep{Pojmi02}; 
(3) \citet{Rinner03}; (4) \citet{Hubscher10}; (5) \citet{Diethelm09}; (6) \citet{Hubscher11}; 
(7) \citet{Diethelm10}; (8) \citet{Nelson11}; (9) \citet{Diethelm11}; (10) \citet{Hubscher12}; (11) \citet{Diethelm12}; 
(12) \citet{Nelson14}; (13) \citet{Hubscher14}; (14) \citet{Nelson15}; (15) \citet{Alton18};   (16) \citet{Hubscher15}; (17) \citet{Hubscher17}; (18) \citet{Nagai19}.
\end{tabnote}
\end{table*}

\begin{figure}[htb]
\begin{center}
   \includegraphics[angle=0,scale=0.6]{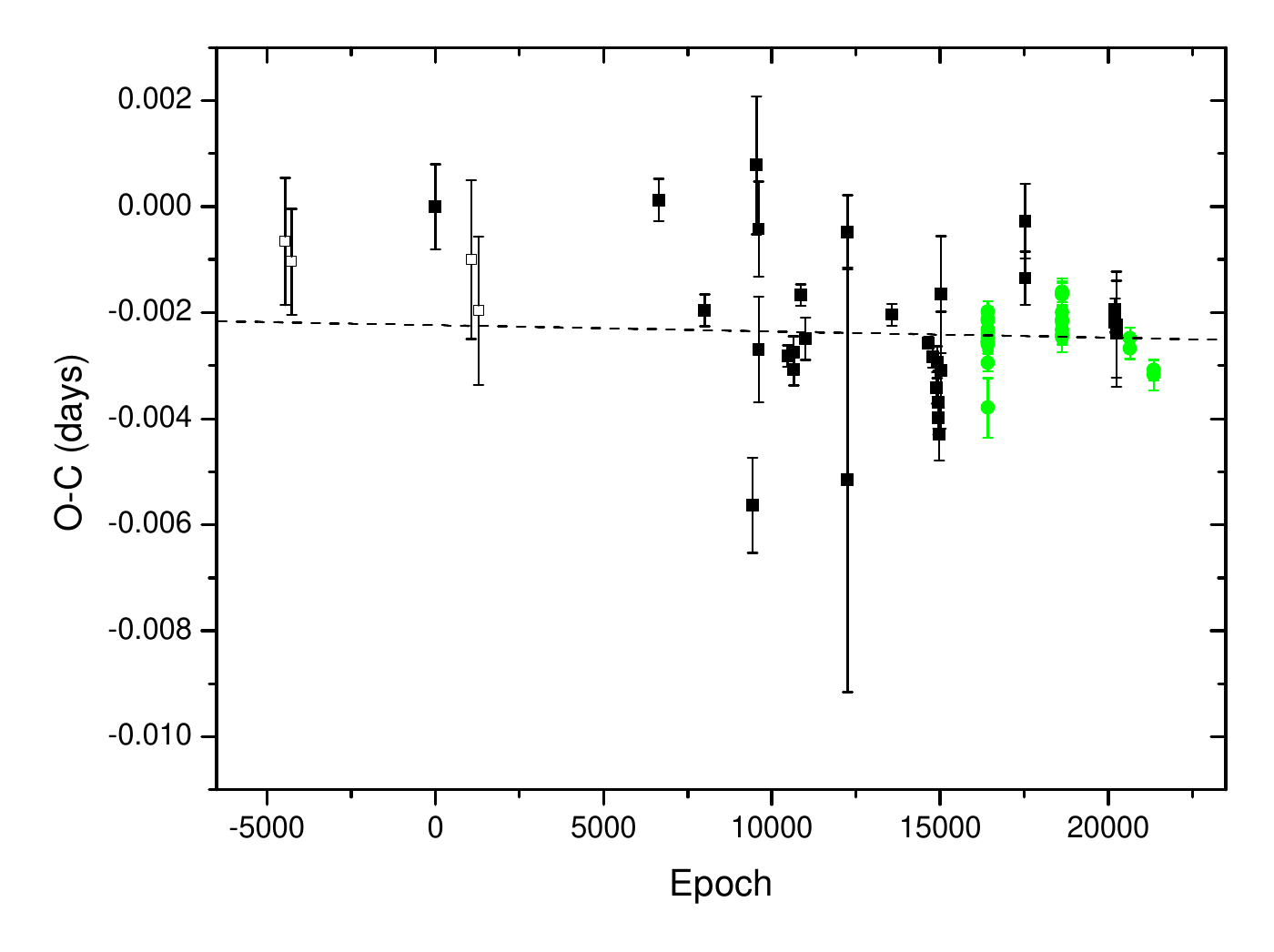}
   \caption{The O$-$C diagram and the result of linear fit (dashed line).
      The green filled circles, black open squares and filled squares denote the O$-$C data 
      newly determined, reprocessed, and from references respectively
   } \label{figocln}
\end{center}
\end{figure}

Combining these values with those got from us (only ground-based result here. For more results see Section \ref{seckep}), the O$-$C diagram was derived and shown in Figure \ref{figocln}. Two data were not used because their O$-$C values are quite different from or in conflict with other neighboring O$-$C data, which makes them unreliable. A linear fit was conducted to the O$-$C diagram and the ephemeris was updated to be
\begin{equation}
\mathrm{min.I~(HJD)} = 2452721.5683(4) + 0^{\mathrm{d}}.26765599(2)\times E
\end{equation}
Further analysis about the O$-$C diagram and the orbital period  was presented in Section \ref{seckoc}, where the times of minimum light from the spacecraft mission Kepler were employed.

\subsection{Photometric solutions}\label{secphot}
We analyzed light curves by using the 2013 version of Wilson-Devinney code \citep{Wilson71,Wilson79,Wilson90,Wilson94,VanH07,Wilson08,Wilson10,Wilson12} (hereafter W-D code). It is a powerful tool for analyzing light curves of eclipsing binaries. This version enables the automatic calculation of limb-darkening coefficients as well as incorporation of aging (grow and decay) spots (use "evolutionary spots" instead in the following). Detailed information about the code could be found in its manual\footnote{ftp://ftp.astro.ufl.edu/pub/wilson/}.

\begin{figure}[t]
\begin{center}
\includegraphics[angle=0,width=8.3cm]{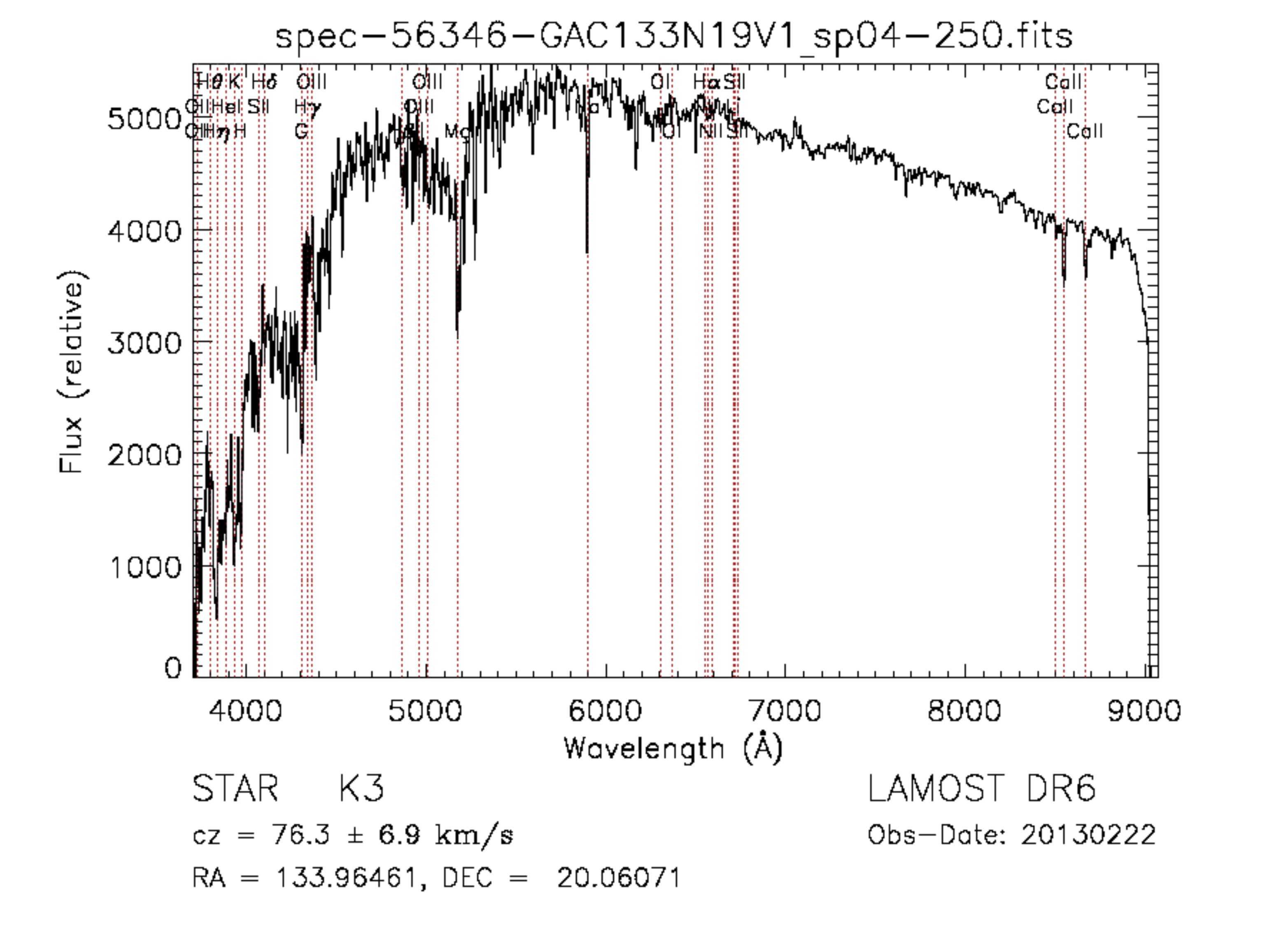}
\caption{
The spectrum of IL Cnc released by Lamost DR6 database \citep{Hebo17}}
\label{figspec}
\end{center}
\end{figure}

In order to start the calculation, we attempted to estimate the temperature of the component stars. According to the color indices $B-V=0.955$ and $J-H=0.452$ of IL Cnc from vizier database\footnote{http://vizier.u-strasbg.fr/viz-bin/VizieR, operated at CDS, Strasbourg, France\label{ftvizier}}, a spectral type of K0-K2 was estimated \citep{Cox00}. Thus, the temperature of star 1 (the component eclipsed at min I) was set to be $T_1 = 5000 $ K ($T_1$ is usually a little bit higher than the average). This temperature indicates convective envelope for the components. Accordingly, the gravity-darkening coefficients were set $g_1 = g_2 = 0.32$ \citep{Lucy67} and the bolometric albedo $A_1 = A_2 = 0.5$ \citep{Rucini69}. The square-root functions ($LD=-3$) were chosen for the treatment of limb-darkening. The corresponding coefficients were calculated by the code according to Van Hamme's table (\yearcite{VanH93}). It is to mention that low-resolution spectra of this target (one example is shown in Figure \ref{figspec}) were obtained by the team of LAMOST, and a K3 spectral type was suggested. Gaia catalogue\footnote{https://gea.esac.esa.int/archive/} also lists the temperature of IL Cnc to be 4894 K. Thus, the temperature estimation should be plausible.

\begin{figure}[htbp]
\begin{center}
   \includegraphics[angle=0,width=8.1cm]{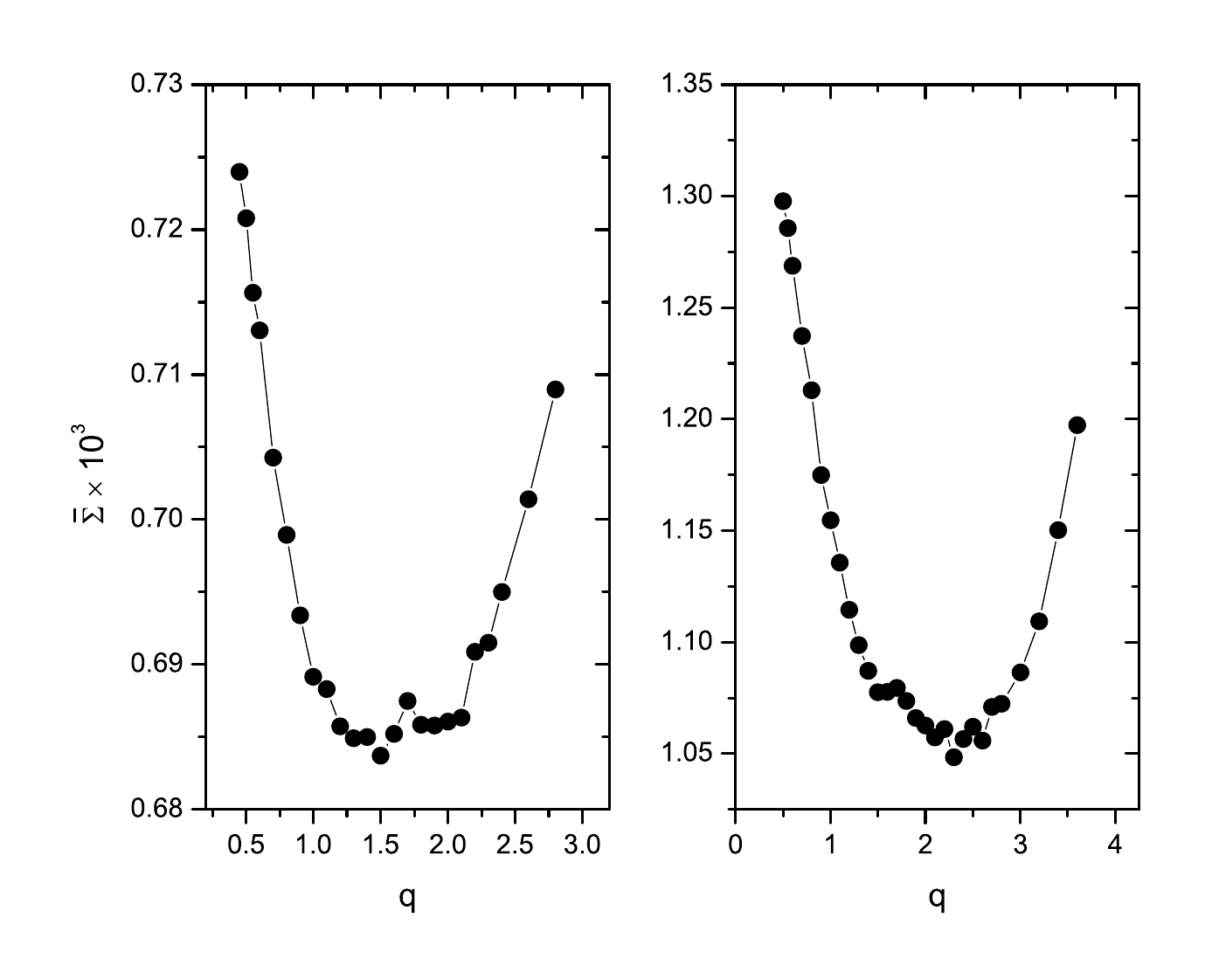}
      \caption{$\overline\Sigma$ - $q$ curves for light curves of ILCnc 
      from 2.4-m telescope in Chiangmai (left) and 70 cm telescope in Lijiang (right)
      } \label{figqser}
\end{center}
\end{figure}

Mode 3 (contact model) was assumed initially for this object according to its EW-type light curves and thus the dimensionless potential of  components, $\Omega_1$ and $\Omega_2$ were bound to be the same. For other configurations (e.g. Mode 2), the potentials should be set according to the Roche potential limits. Some adjustable parameters in the beginning of the calculation were: the orbital inclination $i$; the mean temperature of star 2, $T_2$; the monochromatic luminosity of star 1, $L_{1X}$ (X = B,V,R,I) and the dimensionless potential. The q-search method was applied and solutions were derived for a series of mass ratios. The mean sum of weighted square deviations ($\overline{\sum}$, hereafter mean residuals) along with mass ratios are plotted in Figure \ref{figqser}. For the two sets of light curves, the calculations were started from the same initial parameter values and carried out independently in order to make a comparison.

The minimum of mean residuals were achieved at different values of mass ratio for the two sets ($q=1.5$ and $q=2.3$ respectively) after the q-search. Then, the mass ratios were set free to start calculation for comprehensive solutions. Here comprehensive solutions refer to that starspots, third lights or other assumptions were taken into consideration. These assumptions are widely accepted in the study of late type short period eclipsing binaries. It should be noted that we don't have bias as to whether to add cool or hot spots to the surface of either component in models because we think any situation is possible unless the data demonstrate it.

\begin{figure*}
\begin{center}
   \includegraphics[angle=0,scale=0.85]{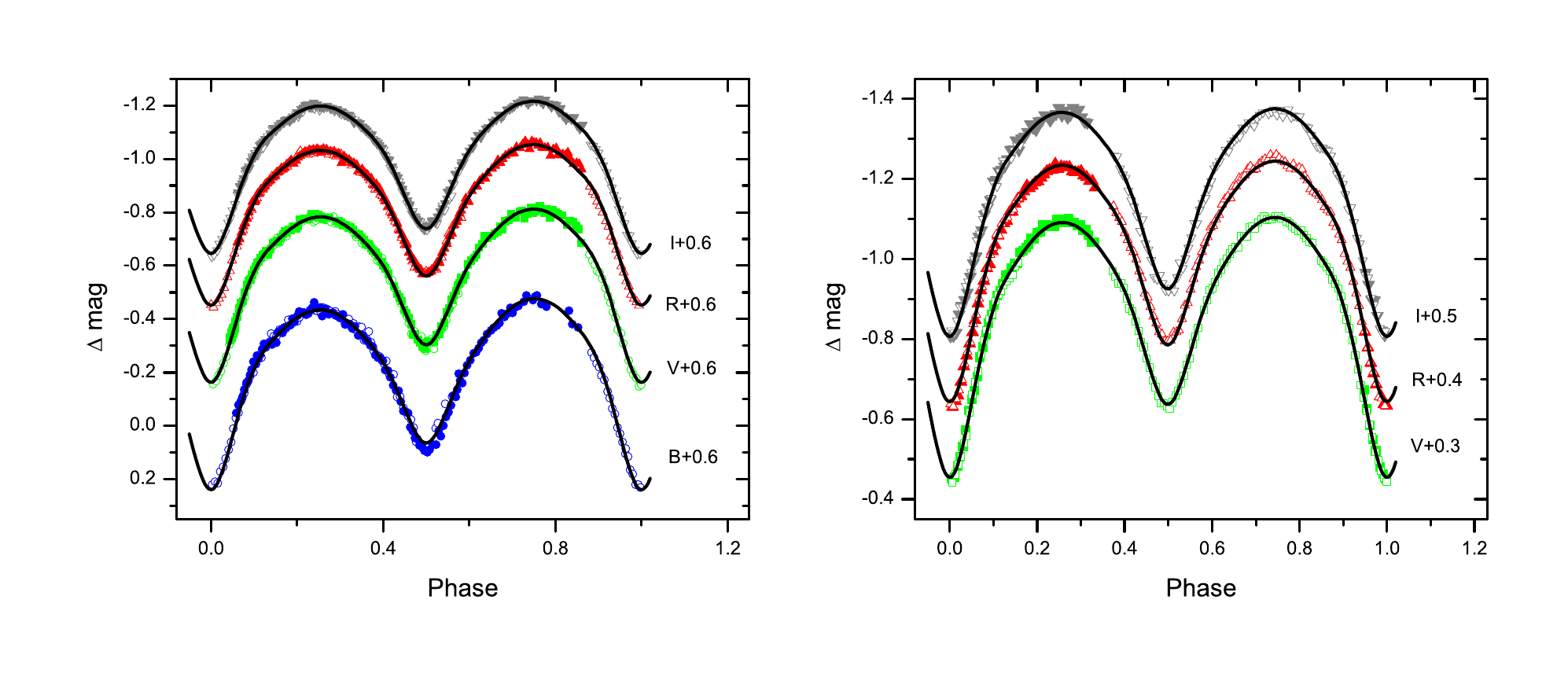}
      \caption{The observed (symbols) and theoretical light curves (solid lines) of IL Cnc. The left and right panels show the results of LCs 2015 and LCs 2016, respectively. The filled and open symbols denote light curves obtained on the first and second night, while the circles, squares, triangles and inverted triangles are for B, V, R and I filters, respectively. The theoretical light curves are calculated from the normal spot solutions in Table \ref{tabwd}}\label{figwdsp}
\end{center}
\end{figure*}

\begin{figure*}
\begin{center}
   \includegraphics[angle=0,width=12.0cm]{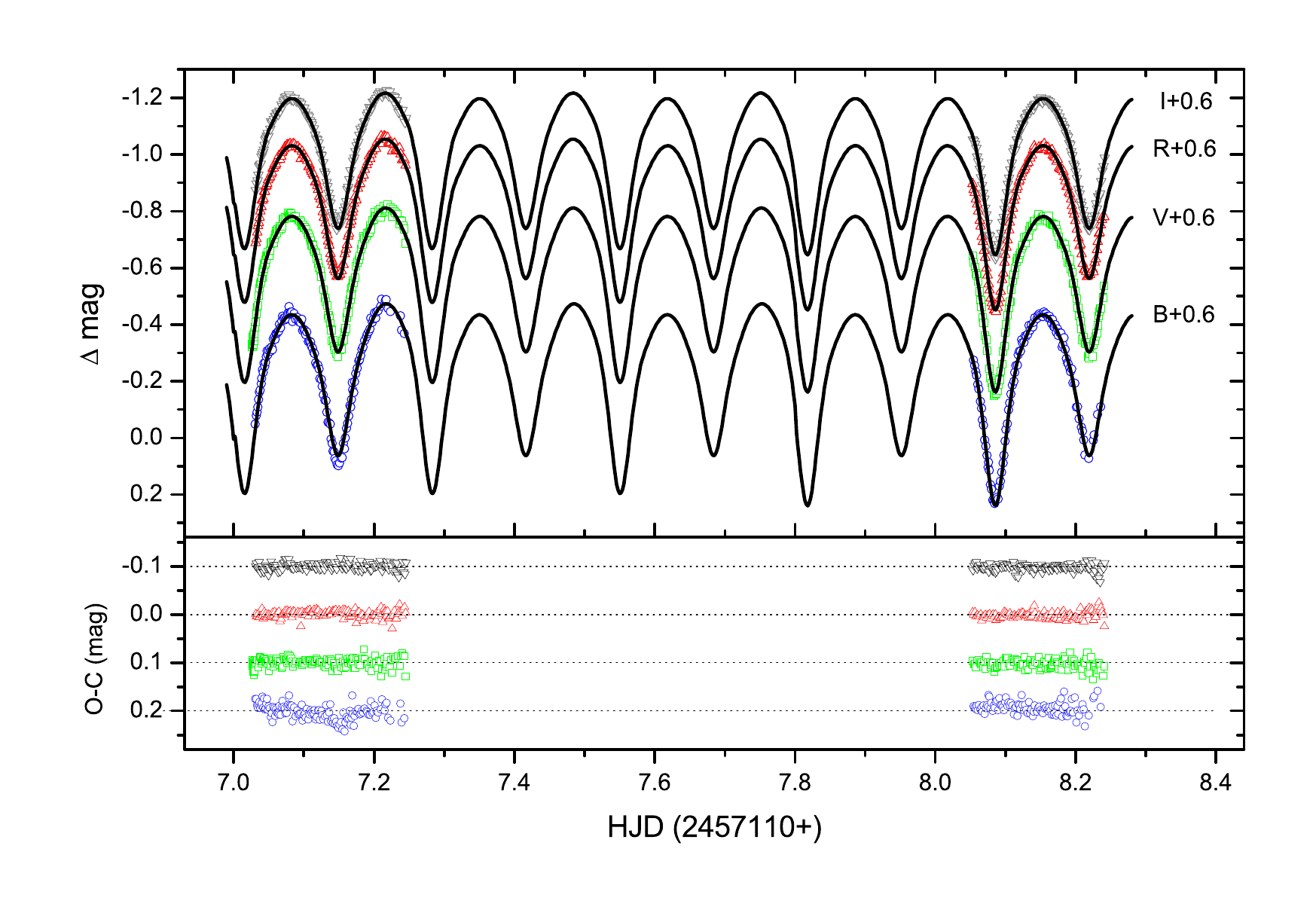}
      \caption{The observed (symbols) and theoretical light curves (solid lines) of LCs 2015 together with the residuals (lower parts). 
The open circles, squares, triangles and inverted triangles are for B, V, R and I filters, respectively. The theoretical light curves are calculated from the ``e-spot'' solutions in Table \ref{tabwd}}\label{figwdesp}
\end{center}
\end{figure*}

The process of calculation was complicated and time consuming while we brought in a grid-search method (see Chapter 8 in the book of \citet{Bevington03}) in seeking for the best solutions. The final solutions are listed in Table \ref{tabwd} for each main case. Parameters for spots are $\theta_{s}$ (latitude), $\psi_{s}$ (longitude), $r_{s}$ (radius) and $T_{s}/T_{\ast}$ (temperature factor), as well as which component they locate on. The parameters which we attempted griding values are marked "trial". "LCs 2015" and "LCs 2016" in the table denote the light curve data from Chiangmai 2.4-m telescope and Lijiang 70cm telescope respectively. It should be noted that for LCs 2015, we tried an evolutionary starspot, which is a new function of W-D code 2013 as mentioned above. The reason we attempted the evolutionary starspot is that we found the light curves have slightly changed during the two continuous nights. This kind of solutions is marked ``e-spot'' in Table \ref{tabwd}. We assume that ``e-spot'' was seen in the system for only one night and then found the best solutions with ``e-spot'' present in the first night. This assumption may be somewhat arbitrary, so this set of solutions might be a tentative result. The derived light curves of the best solutions with normal spots (column 2 and 5) and ``e-spot'' are shown in Figures \ref{figwdsp} and \ref{figwdesp}, respectively. The $\chi^2_{v}$ of each set of solutions are also listed. In order to calculate this parameter, the errors of the light curve data are needed. Combining the errors of C$-$Ch data, those errors were estimated to be 0.012 mag (B), 0.011 mag (V), 0.009 mag (R) and 0.009 mag (I) for LCs 2015, and 0.006 mag (V), 0.005 mag (R),and 0.009 mag (I) for LCs 2016. The $\chi^2_{v}$ and $\overline\Sigma$ values imply that for both sets of light curves, solutions with spots are significantly better than solutions without them. It should be mentioned that we tried adding a third light in our calculation but no reasonable third lights were achieved, which suggests the third light should be less than 2-3$\%$ of the total luminosity, taking account of the errors.

\begin{table*}
\caption{Photometric solutions for the light curves of IL Cnc.}\label{tabwd}
\small
\begin{center}
\begin{tabular}{@{}lll|llll@{}}
\hline %%\tableline
Parameters              & \multicolumn{2}{c|}{LCs 2016} & \multicolumn{4}{c}{LCs 2015}\\
                        & No spots & With spots & No spots  & cool spots & hot spots & e-spots\\
\hline %%\tableline
$q$ ($M_2/M_1$)         & 2.478(7)   & 1.812(6)   & 1.657(5)   & 1.530(6)   & 1.949(5)   & 1.990(5)   \\
$\Omega_{in}$           & 5.9148     & 4.9829     & 4.7585     & 4.5702     & 5.1787     & 5.2371     \\
$\Omega_{out}$          & 5.3054     & 4.3921     & 4.1738     & 3.9913     & 4.5832     & 4.6403     \\
$T_{1}$ (K)             & 5000$^a$   & 5000$^a$   & 5000$^a$   & 5000$^a$   & 5000$^a$   & 5000$^a$   \\
$T_{2}$ (K)             & 4633(5)    & 4709(8)    & 4729(4)    & 4735(3)    & 4727(4)    & 4731(3)    \\
$i(^{\circ})$           & 73.64(10)  & 73.32(8)   & 73.67(8)   & 73.93(6)   & 73.84(5)   & 73.97(5)   \\
$L_{1}/L_{total}$ (B)   & ---        & ---        & 0.495(1)   & 0.511(1)   & 0.459(1)   & 0.452(1)   \\
$L_{1}/L_{total}$ (V)   & 0.418(1)   & 0.459(1)   & 0.473(1)   & 0.490(1)   & 0.437(1)   & 0.431(1)   \\
$L_{1}/L_{total}$ (R)   & 0.394(1)   & 0.440(1)   & 0.455(1)   & 0.472(1)   & 0.420(1)   & 0.414(1)   \\
$L_{1}/L_{total}$ (I)   & 0.379(1)   & 0.428(1)   & 0.444(1)   & 0.462(1)   & 0.409(1)   & 0.403(1)   \\
$\Omega_{1}=\Omega_{2}$ & 5.822(13)  & 4.934(10)  & 4.699(6)   & 4.486(8)   & 5.122(7)   & 5.184(8)   \\
$r_{1}$(pole)           & 0.2905(6)  & 0.3120(5)  & 0.3204(3)  & 0.3297(3)  & 0.3068(3)  & 0.3048(3)  \\
$r_{1}$(side)           & 0.3039(7)  & 0.3265(6)  & 0.3358(4)  & 0.3461(3)  & 0.3210(3)  & 0.3188(4)  \\
$r_{1}$(back)           & 0.3418(10) & 0.3616(8)  & 0.3716(6)  & 0.3834(5)  & 0.3567(5)  & 0.3544(6)  \\
$r_{2}$(pole)           & 0.4389(11) & 0.4103(11) & 0.4042(8)  & 0.4002(11) & 0.4169(8)  & 0.4183(9)  \\
$r_{2}$(side)           & 0.4700(16) & 0.4358(15) & 0.4289(11) & 0.4246(15) & 0.4436(11) & 0.4452(11) \\
$r_{2}$(back)           & 0.4994(21) & 0.4661(21) & 0.4602(15) & 0.4576(23) & 0.4737(15) & 0.4751(16) \\
$f(\%)$                 & 15.2(2.1)  & 8.1(1.7)   & 10.2(1.1)  & 14.5(1.3)  & 9.5(1.2)   & 8.9(1.3)   \\
$\theta_{s}(^{\circ})$  & ---        & 34(trial)  & ---        & 134(trial) & 141(trial) & 63(trial)  \\
$\psi_{s}(^{\circ})$    & ---        & 331(4)     & ---        &  92(2)     & 256(2)     & 233(2)     \\
$r_{s}(^{\circ})$       & ---        & 24.0(1.0)  & ---        &  19.2(4)   & 15.2(3)    & 21.2(4)    \\
$T_{s}/T_{\ast}$        & ---        & 0.92(trial)& ---        & 0.65(trial)& 1.30(trial)& 1.06(trial)\\
Spots on star           & ---        & 1          & ---        & 2          & 2          & 2          \\
$\overline\Sigma \times 10^3$ & 1.041 & 0.899     & 0.685      & 0.557      & 0.538      & 0.496      \\
$\chi^2_{v}$            & 3.03       & 1.95       & 1.55       & 1.11       & 1.04       & 0.92       \\
\hline %%\tableline
\end{tabular}
\end{center}
\scriptsize
$^{(a)}$Assumed. $L_{total}=L_{1}+L_{2}$. ``trial'' denotes the parameter
was fixed at a series of trial values in the calculation until the best value was found.
\end{table*}

\section{Anlysis of Kepler data}\label{seckep}
The Kepler spacecraft and telescope were designed on the main purpose of searching for exoplanets, which however has provided a large amount of high precise and long-term monitored photometric data for eclipsing binaries. IL Cnc was one of them that benefited from this mission in its rejuvenated stage - K2 \citep{Howell14}. The target was observed in the Kepler K2 mission (hereafter K2) with long cadence mode which took 30 minutes for each image. It was monitored during Kepler day 2307 to 2382 and 3419 to 3470 which corresponds to campaign 5 and campaign 18. The data were downloaded from Mikulski Archive for Space Telescopes (MAST) archive in the form of two fits files, and the light curves (hereafter ``lc05" and `lc18") were extracted then. The light curves (only a part of lc05 as an example) vs BJD (barycentric Julian Date) and the phased light curves (rebinned, see Section \ref{seckphot}) are shown in Figure \ref{figlckp}. The light curves used for analysis are PDCsap data. We did not perform further detrending procedure because the general trend of them are flat and detrending may cause further problems.

\subsection{Information from maximum and minimum light}\label{seckoc}
\begin{figure}
\begin{center}
   \includegraphics[angle=0,scale=0.6]{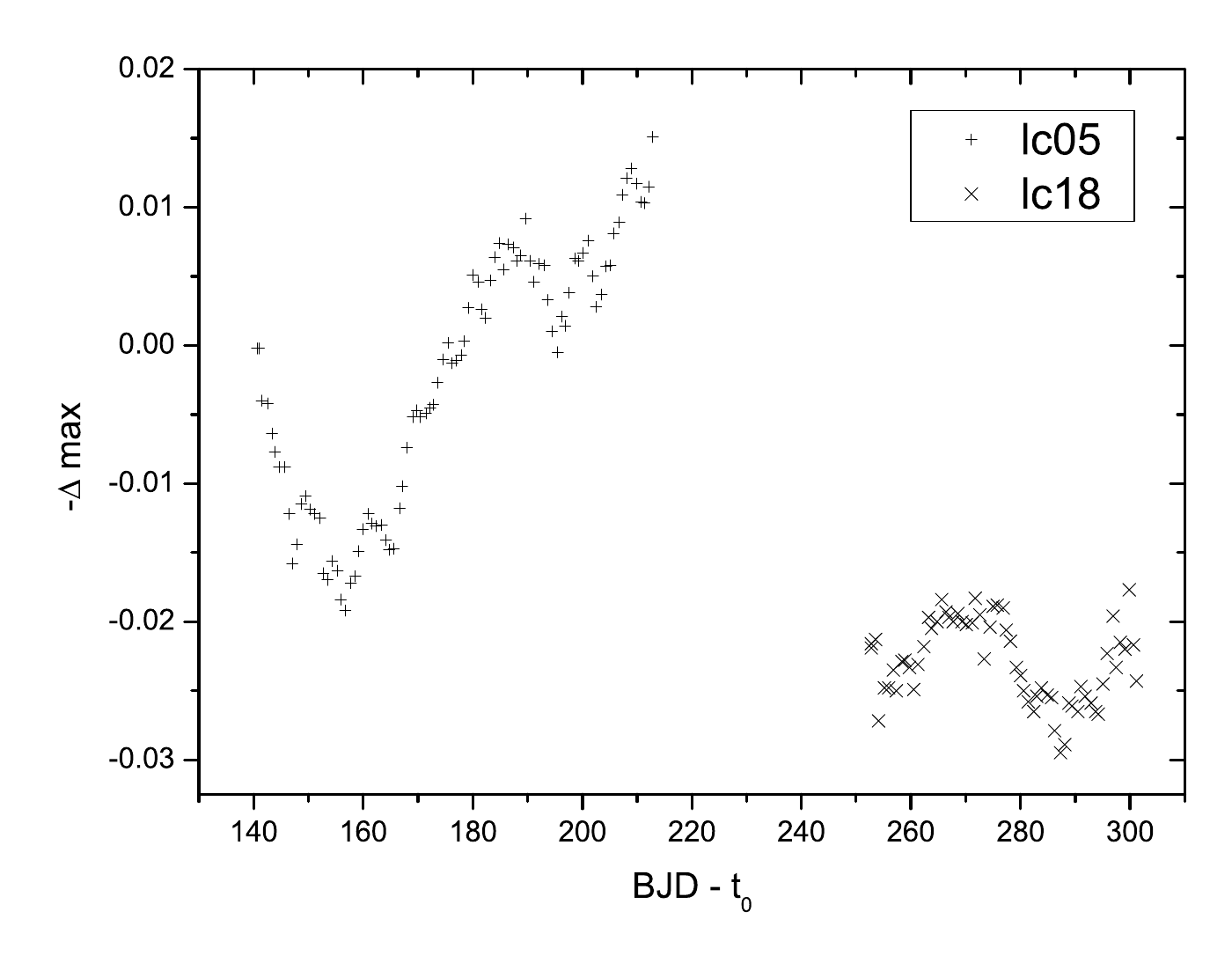}
      \caption{The O'Connell effect of IL Cnc from K2 data. Here $\textrm{t}_0$ equals to 2457000 and 2458000 for lc05 and lc18, respectively.}\label{figconn}
\end{center}
\end{figure}

From Section \ref{secphot}, it is known that O'Connell effects present in the light curves of this target. To our expectation, this effect was also found prominent in K2 data. It was analyzed by calculating the difference of two maxima of the light curves in each cycle, which is shown in Figure \ref{figconn}. The maxima were determined with a least squares parabolic fitting method to the combined data (explained in the following paragraph). It is shown clearly that the O'Connell effect changes with time. The shape of the light curve might be changing all the time, which demonstrates that the target is highly active, and so that the ``e-spot'' scenario in Section \ref{secphot} is plausible.

\begin{figure*}
\begin{center}
   \includegraphics[angle=0,width=13cm]{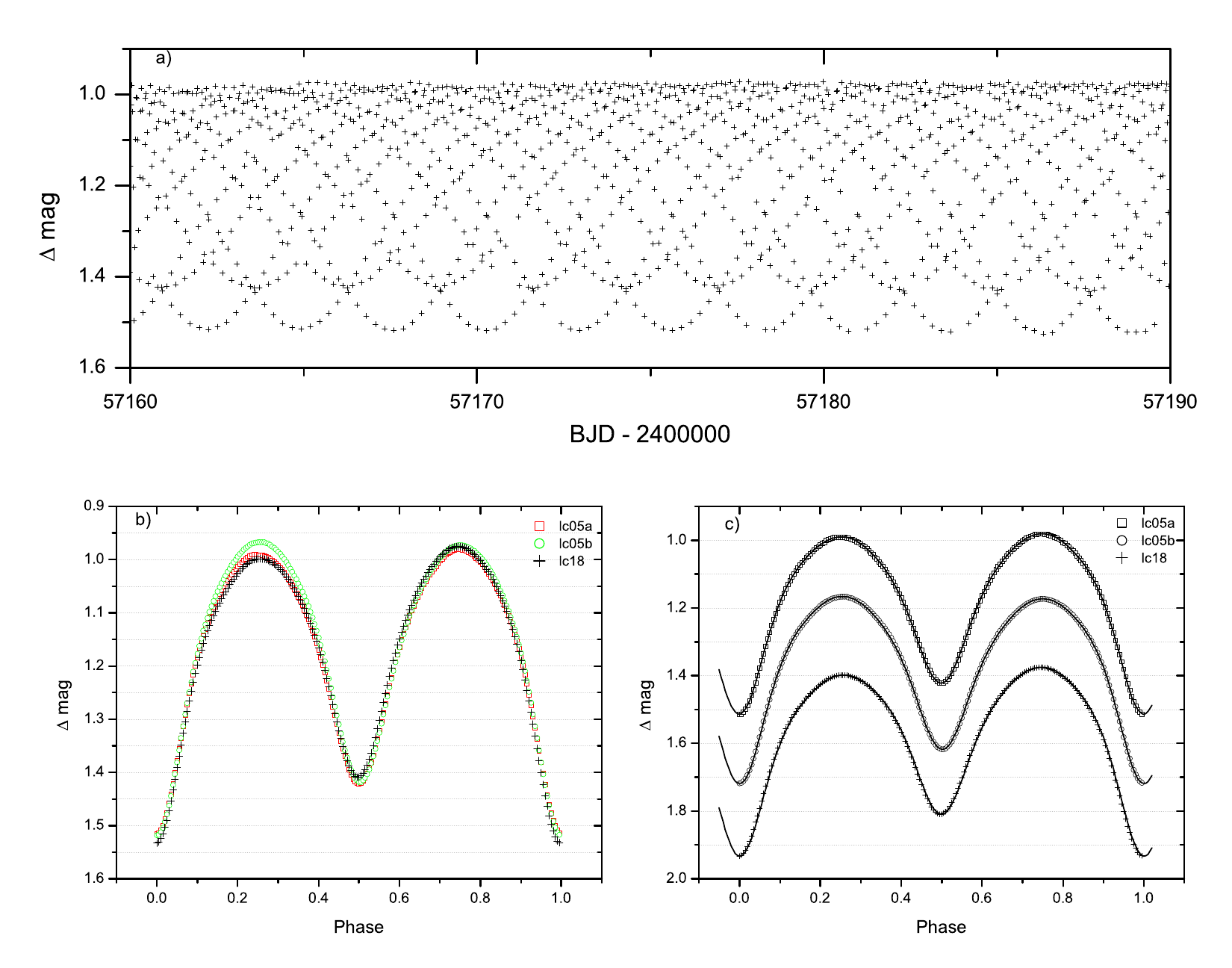}
      \caption{The K2 light curves of IL Cnc. Upper pannel: the light curves vs BJD (only one part of lc05). Lower left: comparison of phased light curves (rebinned) of different segments: lc05a, lc05b and lc18. lower right: same with panel (b) but fitted with the calculated light curves from best corresponding solutions in Table \ref{tabkpwd}. The light curves in panel (c) are vertically shifted for better visualization. }\label{figlckp}
\end{center}
\end{figure*}

\begin{figure}[htb]
\begin{center}
   \includegraphics[angle=0,scale=0.63]{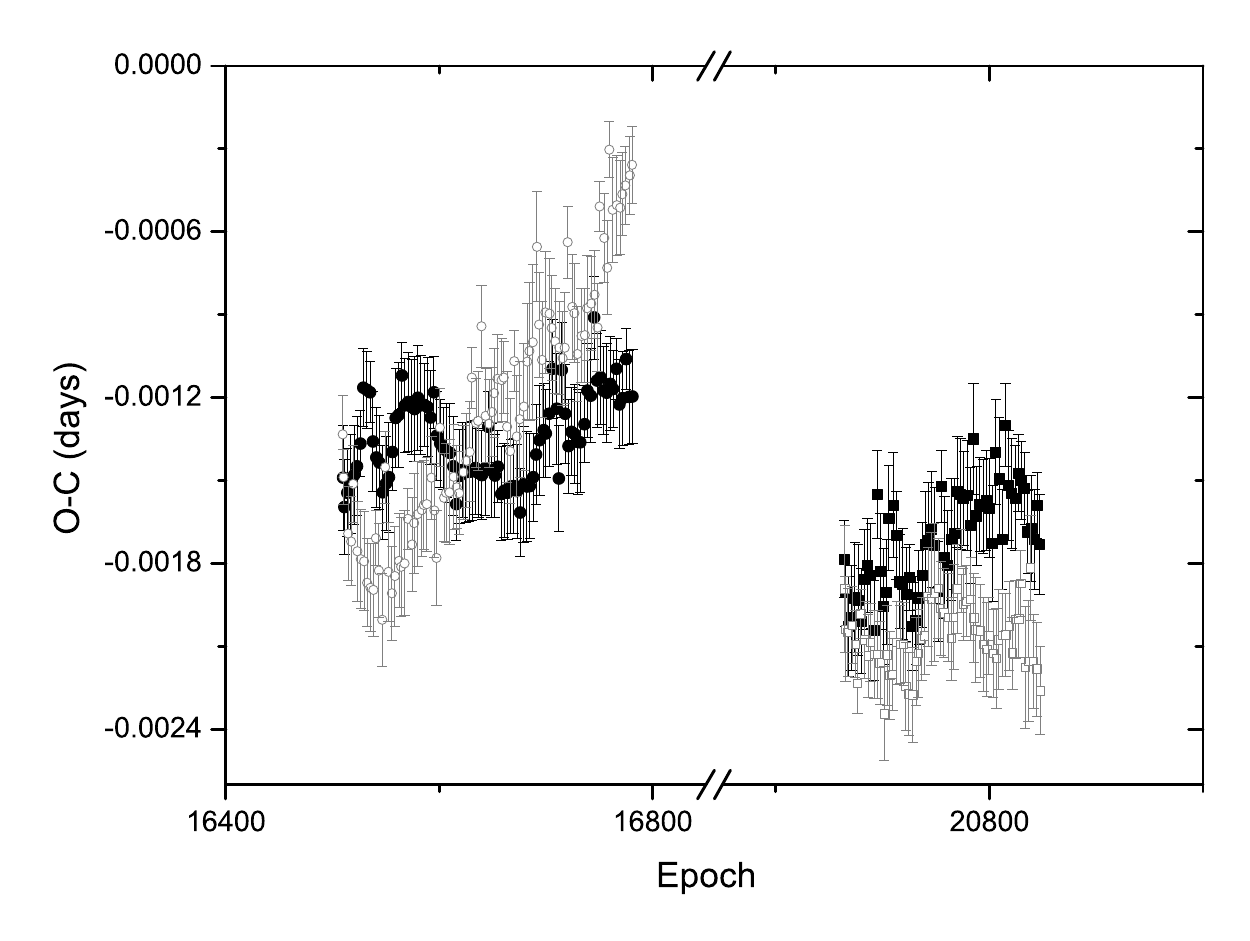}
   \caption{The O$-$C diagram of K2 data. The circles and squares denote data from lc05 and lc18, respectively, while filled black and open grey symbols represent primary and secondary minima, respectively. 
   }\label{figockc}
\end{center}
\end{figure}

\begin{figure}[htbp]
\begin{center}
   \includegraphics[angle=0,scale=0.63]{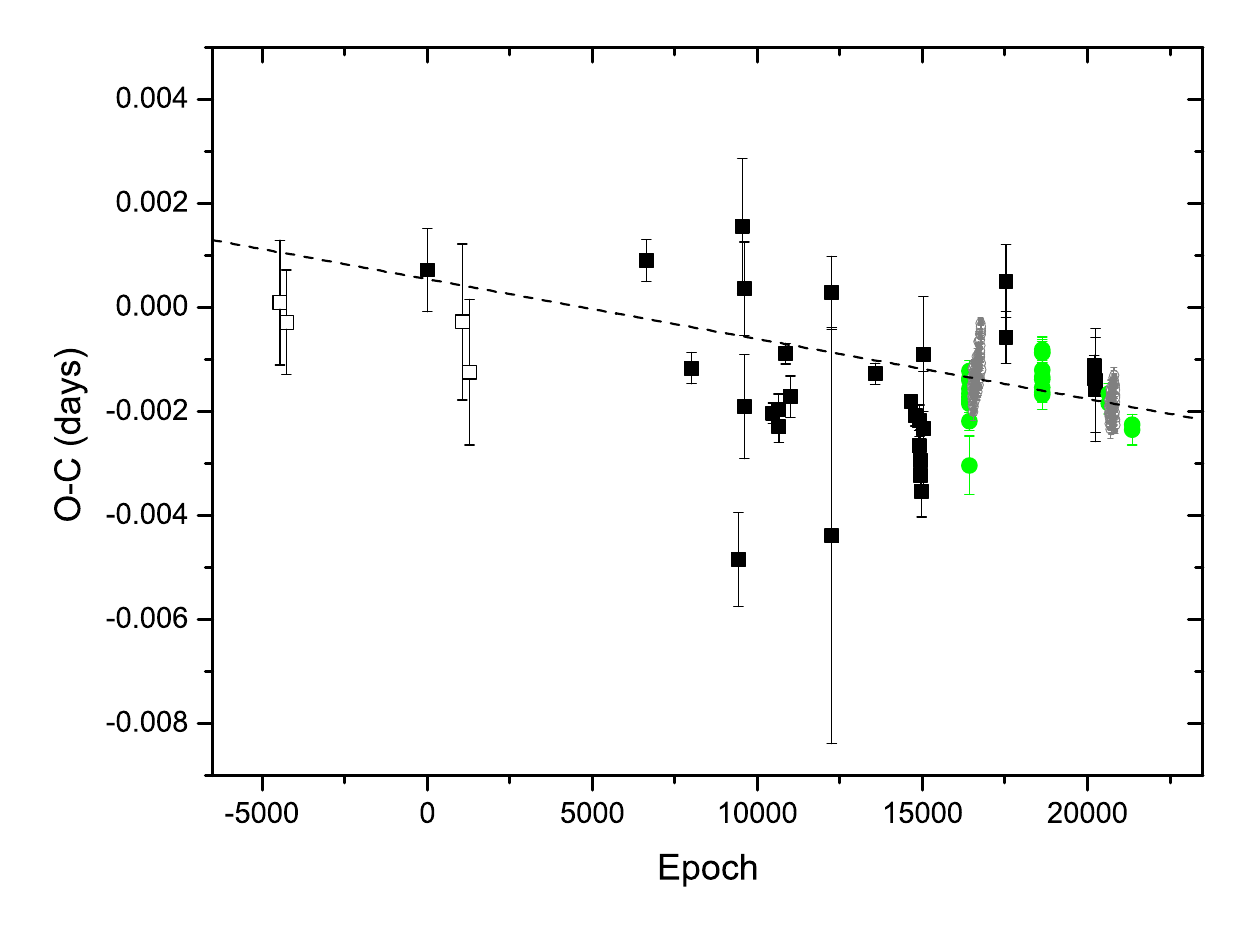}
   \caption{The O$-$C diagram with K2 data and the new linear fit (dashed line). The open grey circles represent the data from K2 while others symbols have the same meaning as they are in Figure \ref{figocln}.
   } \label{figocaln}
\end{center}
\end{figure}

All the minimum light times determined from K2 data are listed in Table \ref{tabkpmin}. Since there are much less from enough data in each cycle of the light curves, we tried to use the following combining method to determine minimum light times. First, combine data in 10 continuous cycles and fold them in period to determine one minimum (maximum) data, and then shift 3 cycles to determine another one. When meeting the interrupted cycles (or the starting and ending parts), make sure there are at least 7 cycles for one measurement. This method helps to ensure there are enough data to determine one data point and avoid large spanning of time. The O$-$C data are calculated using the same formula as used in Section \ref{secper}. It should be noted that since the timings of K2 data are in BJD, other data are then unified into BJD\footnote{http://astroutils.astronomy.ohio-state.edu/time/hjd2bjd.html} as well for the following analysis. The O$-$C diagram for K2 data is shown in Figure \ref{figockc}. The overall O$-$C diagram was replotted and shown in Figure \ref{figocaln}. The new ephemeris was derived to be
\begin{equation}
\mathrm{min. I~(BJD)} = 2452721.5699(2) + 0^{\mathrm{d}}.26765588(1)\times E
\end{equation}

\begin{figure}
\begin{center}
   \includegraphics[angle=0,scale=0.63]{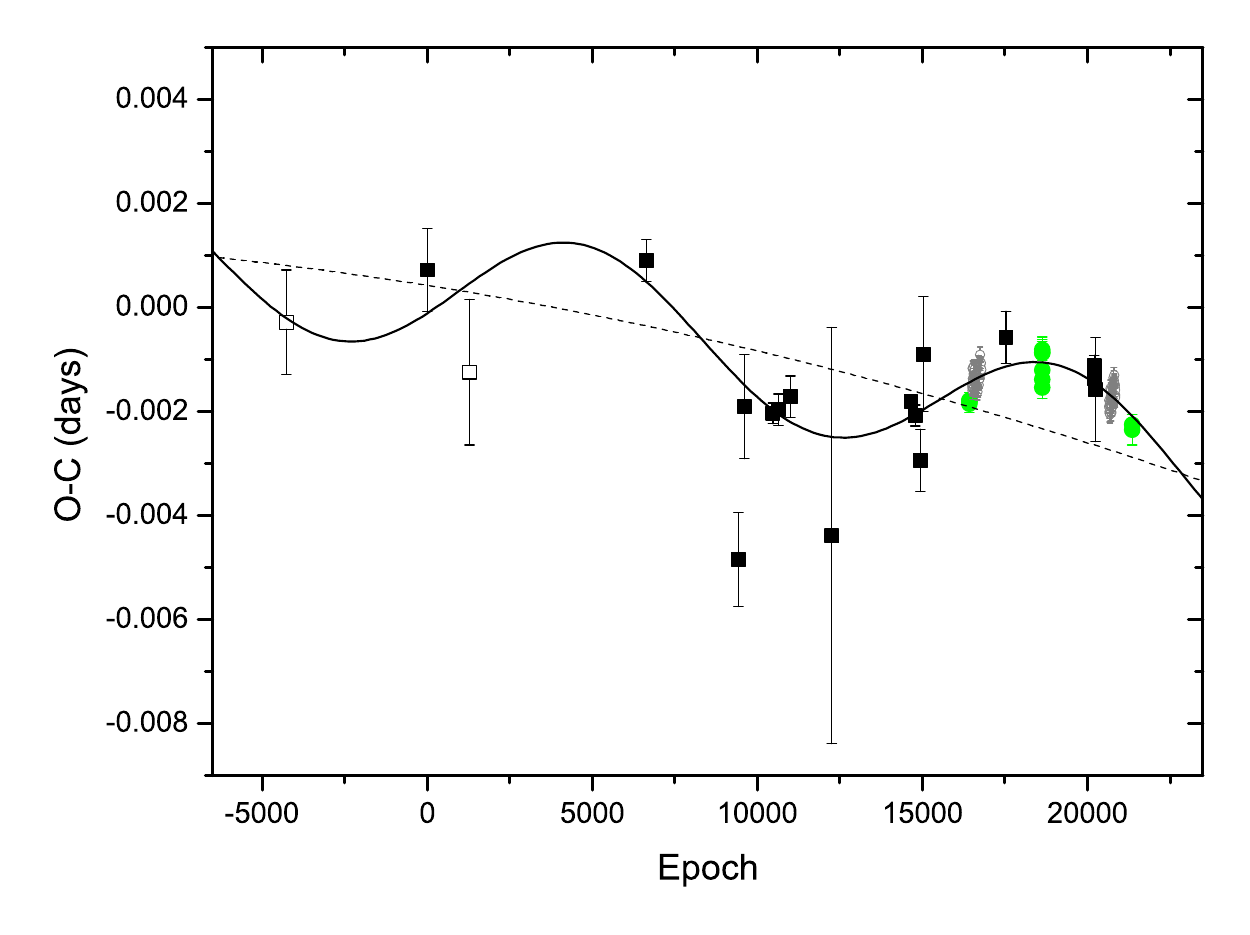}
   \caption{The O$-$C diagram for primary minima. The symbols have the same meaning as they are in Figure \ref{figocaln} but only for primary minima. The solid and dashed line denote the cyclic fit and its quadratic component, respectively.
   }\label{figockpr}
\end{center}
\end{figure}

Seen from Figure \ref{figockc}, both primary and secondary minima change with time, but they vary differently. Comparing Figure \ref{figockc} with Figure \ref{figconn}, it is found that the variation of secondary minima in Figure \ref{figockc} and the O'Connell effect in Figure \ref{figconn} look quite similar. It might be the case that the times of minimum light are shifted by the spot activities as pointed out by many investigators \citep{Kalimeris02, Watson04}. However, because the primary eclipse is deeper than the secondary (see Figures \ref{figlcs} and \ref{figlckp}), 
the primary minima may not be so seriously affected by the spot activities, as shown in Figure \ref{figockc}. Therefore, we tried to analyze the primary minima separately. As shown in Figure \ref{figockpr}, a cyclic variation might be presenting. Using a sinusoidal function for the periodic component, the ephemeris was derived to be
\begin{equation}\label{eq:epocy}
\begin{array}{ll}
\mathrm{min. I~(BJD)} =& 2452721.5709(1) + 0^{\mathrm{d}}.26765590(4)\times E \\
                       & -0^{\mathrm{d}}.25(18)\times10^{-11}\times E^2  \\
                       & + 0^{\mathrm{d}}.00131(13)\times\sin(0^{\circ}.0247(8)\times E - 24(13)^{\circ})
\end{array}
\end{equation}
The resulted curve does fit the data well (expect for a few data with large errors). The reduced chi-squares of the fit was calculated to be $\chi ^2_{v} = 1.35$. The standard deviation of the O-C residuals is only about 0.00020 days, which is significantly smaller than the amplitude ($\sim$ 0.0013 days) of the cyclic component, which may indicate that the cyclic variation is true for the primary minima.

\subsection{Photometric solution of Kepler K2 data}\label{seckphot}
The continuous light curve is pretty useful for studying the photometric properties of the target. However, since the exposure time is 30 minutes, which is rather long compared with its period $\sim 0.267$ days, it is necessary to merge data in neighboring cycles. The data were phased and superimposed using Formula \ref{eq:ephe}. Then the phased light curve were rebinned and fitted to derive the new appropriate light curves for synthetic analysis. The magnitude of one point is an interpolation of the phased data spanning about 30 minutes (approximately equals to the exposure time) and there are altogether 200 data points in one light curve. It should be noted that because the shape of lc05 changes greatly (the O'Connell effect is almost reverted, see Figure \ref{figconn}), we just simply divided them into two parts, lc05a and lc05b (separates from BJD 2457179.0), approximately corresponding to the negative and positive O'Connell effects in lc05. The rebinned light curves are shown in Figure \ref{figlckp}.

For the long cadence data, there must be problem of average effect for the data points, which will make sharp features disappear and change the depth of the eclipse. This is the so called smear effect. The W-D code uses the parameter ``NGA" to account for this issue and it was set to NGA = 2 in our calculation. For other parameters, similar procedures as mentioned in Section \ref{secphot} were conducted. The best photometric solutions were achieved for each light curve.  The results are shown in Table \ref{tabkpwd}. Since both hot and cool spot scenarios led to good fit for lc05a, both results are listed. While for lc05b and lc18, only cool spot scenarios could give good results ($\overline\Sigma$ of hot spot scenarios are much worse) and hence listed. It is inferred from this table that the spot activities do affect the solutions of the light curves and cause large uncertainties of the mass ratio. However, other parameters are yet consistent with each other, especially the inclinations, which implies the results to be reliable on the whole.

\begin{table*}
\caption{Photometric solutions for Kepler data of IL Cnc.}\label{tabkpwd}
\small
\begin{center}
\begin{tabular}{@{}lllll@{}}
\hline %%\tableline
Parameters              & \multicolumn{1}{c}{lc05a} & \multicolumn{1}{c}{lc05a} & 
                \multicolumn{1}{c}{lc05b} & \multicolumn{1}{c}{lc18}  \\
                        & hot spots & cool spots  & cool spots & cool spots \\
\hline %%\tableline
$q$ ($M_2/M_1$)              &  1.442(1)    &   1.397(1)    &  1.713(2)    &  1.425(1)   \\
$\Omega_{in}$                &  4.4386      &   4.3706      &  4.8404      &  4.4141     \\
$\Omega_{out}$               &  3.8642      &   3.7987      &  4.2534      &  3.8406     \\
$T_{2}$ (K)                  &  4756(1)     &   4762(1)     &  4736(2)     &  4711(1)    \\
$i(^{\circ})$                &  73.01(1)    &   73.25(1)    &  73.87(2)    &  73.34(1)   \\
$L_{1}/L_{total} (\%)$       &   48.4(1)    &    49.0(1)    &   45.2(1)    &   50.0(1)   \\
$\Omega_{1}=\Omega_{2}$      &  4.402(1)    &   4.325(1)    &  4.778(2)    &  4.368(1)   \\
$r_{1}$(pole)                &  0.3297(1)   &   0.3333(1)   &  0.3179(1)   &  0.3317(1)  \\
$r_{1}$(side)                &  0.3456(1)   &   0.3497(1)   &  0.3331(1)   &  0.3479(1)  \\
$r_{1}$(back)                &  0.3797(1)   &   0.3843(1)   &  0.3690(2)   &  0.3826(1)  \\
$r_{2}$(pole)                &  0.3906(1)   &   0.3889(1)   &  0.4071(3)   &  0.3906(1)  \\
$r_{2}$(side)                &  0.4130(2)   &   0.4113(2)   &  0.4323(4)   &  0.4133(1)  \\
$r_{2}$(back)                &  0.4442(2)   &   0.4431(2)   &  0.4634(5)   &  0.4450(2)  \\
$f(\%)$                      &  6.3(2)      &   8.0(2)      &  10.6(4)     &  8.1(1)     \\
$\theta_{s}(^{\circ})$       &  125$^{t}$   &   119$^{t}$   &  99$^{t}$    &  112$^{t}$  \\
$\psi_{s}(^{\circ})$         &  76(1)       &   79(1)       &  328(2)      &  117(1)     \\
$r_{s}(^{\circ})$            &  9.4(1)      &   10.4(1)     &  10.0(1)     &  18.9(1)    \\
$T_{s}/T_{\ast}$             &  1.21$^{t}$  &   0.78$^{t}$  &  0.86$^{t}$  &  0.85$^{t}$ \\
Spots on star                &  1           &   2           &  2           &  2          \\
$\overline\Sigma\times 10^3$ &  0.0184      &   0.0201      &  0.0260      &  0.0137     \\
\hline %%\tableline
\end{tabular}
\end{center}
\scriptsize
$^{t}$ ``trial" values, see Table \ref{tabwd}.
\end{table*}

\section{Analysis of Lamost msp data}\label{secmsp}
Except for the Kepler data, the target was also found to be in the field of Lamost survey. This project uses a telescope with effective aperture of 4 meters and a total of 4000 fibers \citep{Liux15}, which makes it a powerful tool for spectral acquisition. Our target was observed by the telescope with both low and median resolution modes. Here, we use median resolution spectra \citep{Wang19} to study for more detailed information, e.g. the radial velocities (RVs). The detailed information about the spectrograph and the survey could be found in the literature (\cite{CuiX12,Zhao12}).

\begin{figure}
\begin{center}
   \includegraphics[angle=0,scale=0.6]{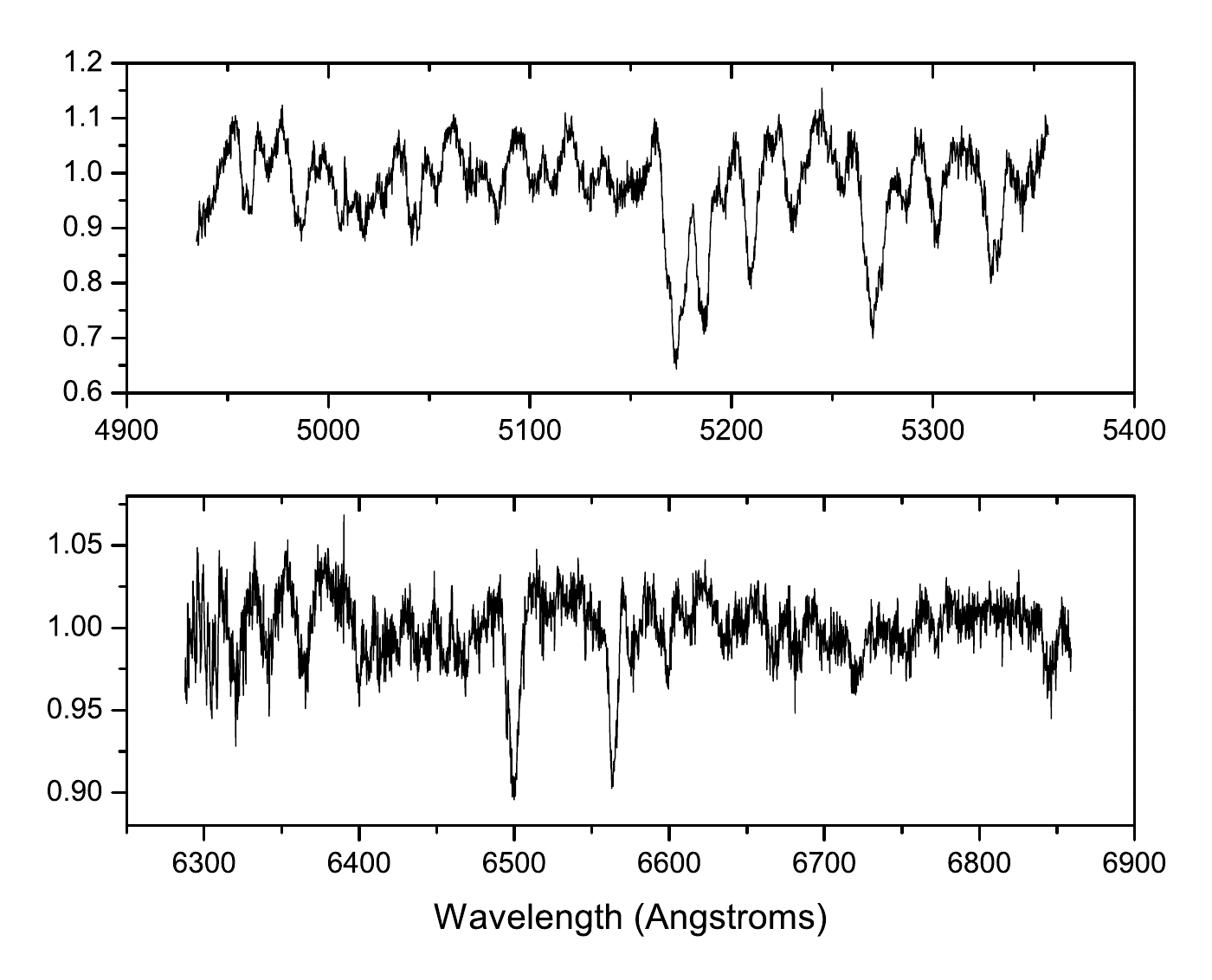}
      \caption{The normalized spectrum of msp119.}\label{fignorm}
\end{center}
\end{figure}

There are six coadding spectra of median resolution that were obtained by Lamost group. However, only three of them are good enough to be used (others have much lower SN ratios). They (hereafter ``msp091'',``msp119'' and ``msp151'') were observed on nights with local MJD 58091, 58119, 58151, with 10 minutes for each single exposure. Each spectrum is a coadding of three spectra observed at almost the same time. They cover two wavelength ranges: 495-535 nm and 630-680 nm which are the so-called blue band and red band \citep{Zong18, Wang19}. Since the spectra were reduced, they were to be used directly. Before analysis, the spectra were normalized using Chebyshev function provided in the astropy package. An example of the normalized spectra is shown in Figure \ref{fignorm}.

\begin{figure}
\begin{center}
   \includegraphics[angle=0,scale=0.6]{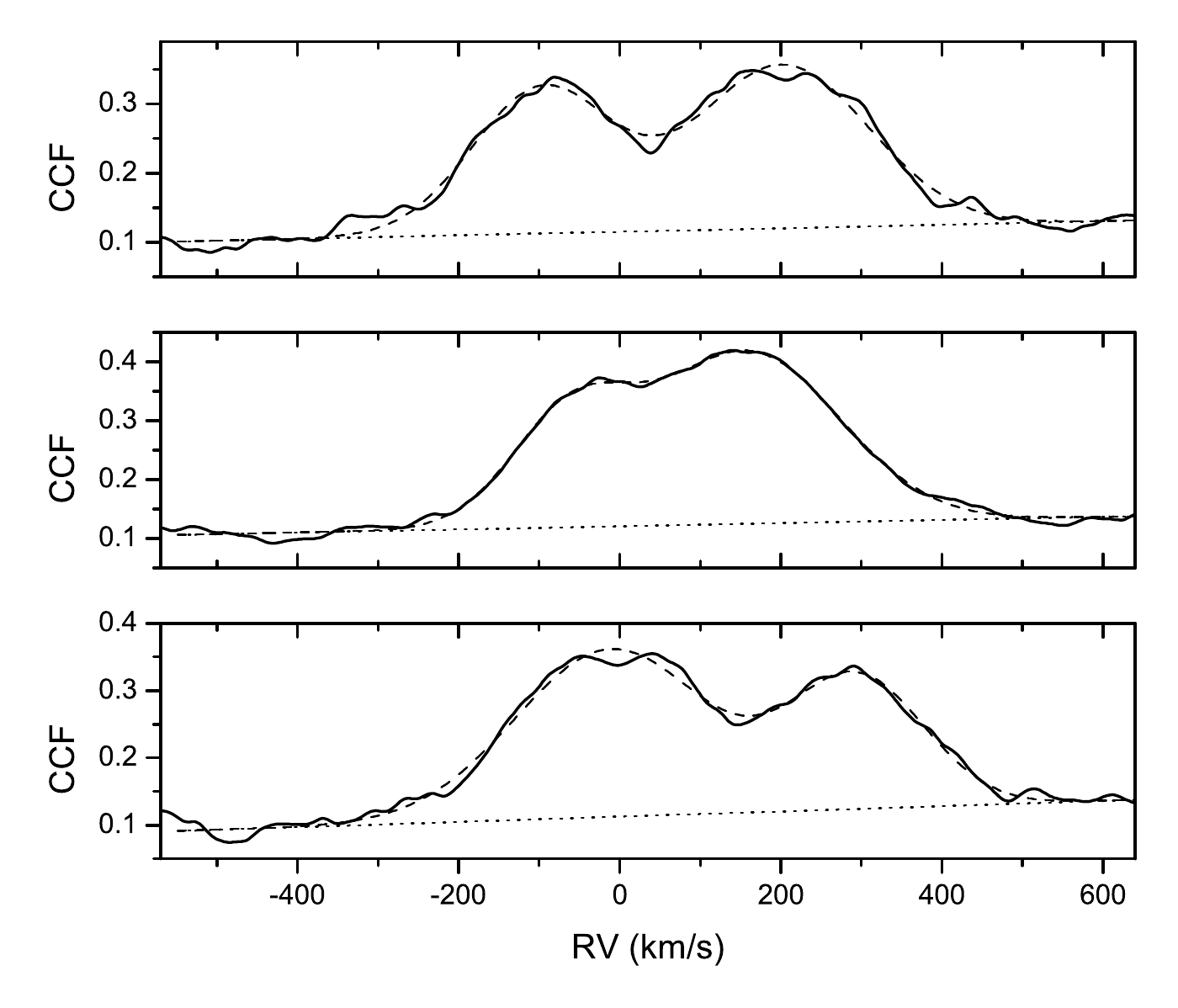}
      \caption{The CCF profile and double gaussian fit. From top to bottom are for msp091, msp119 and msp151, respectively. Solid lines denote the CCF profile. Dashed lines denote the double gaussian fit while dotted lines show the linear component.}\label{figccf}
\end{center}
\end{figure}

To get the RVs, the object spectra were matched with a template spectrum (spectral type K) to obtain CCF (cross correlation function) profiles and then applied double gaussian fit to the profiles. Only spectra in the blue band were used for CCF because the strong absorption of H$\alpha$ in the red band may be easily affected by spot activities. The resulted CCF profiles are shown in Figure \ref{figccf} and the fitting results are listed in Table \ref{tabsplog}, together with a brief log of observations (from the header of fits files). It should be noted that the errors in this table are only the errors of gaussian fit, so the real errors should be larger.

\begin{table*}[htbp]
\caption{The spectroscopy log of IL Cnc.}\label{tabsplog}
\begin{center}
\begin{tabular}{lccccccc}
\hline
Spectra  & DATE-OBS       &  HJD &   phase$^{\ast}$  &   SNR &  SNR &  $V_{r,1}$   &  $V_{r,2}$     \\
      & (UTC)  & (2,400,000+) &          & B band &  R band & (km/s)  & (km/s) \\\hline
msp091  & 2017-12-03 19:50:12.3 &  58091.324268   &   0.1555  &   29  &  58  & $ -96.3\pm 1.4$ & $ 201.3\pm 1.4$   \\
msp119  & 2017-12-31 17:18:27.7 &  58119.221079   &   0.3818  &   50  &  86  & $ -60.2\pm 1.5$ & $ 158.5\pm 1.5$   \\
msp151  & 2018-02-01 15:33:02.3 &  58151.150363   &   0.6741  &   38  &  66  & $ 294.2\pm 1.3$ & $ -10.7\pm 1.2$   \\
\hline
\end{tabular}
\end{center}
\begin{tabnote}
$^{\ast}$ Determined using the first ephemeris in Section \ref{secobs}.
\end{tabnote}
\end{table*}

\begin{figure}
\begin{center}
\includegraphics[angle=0,scale=0.63]{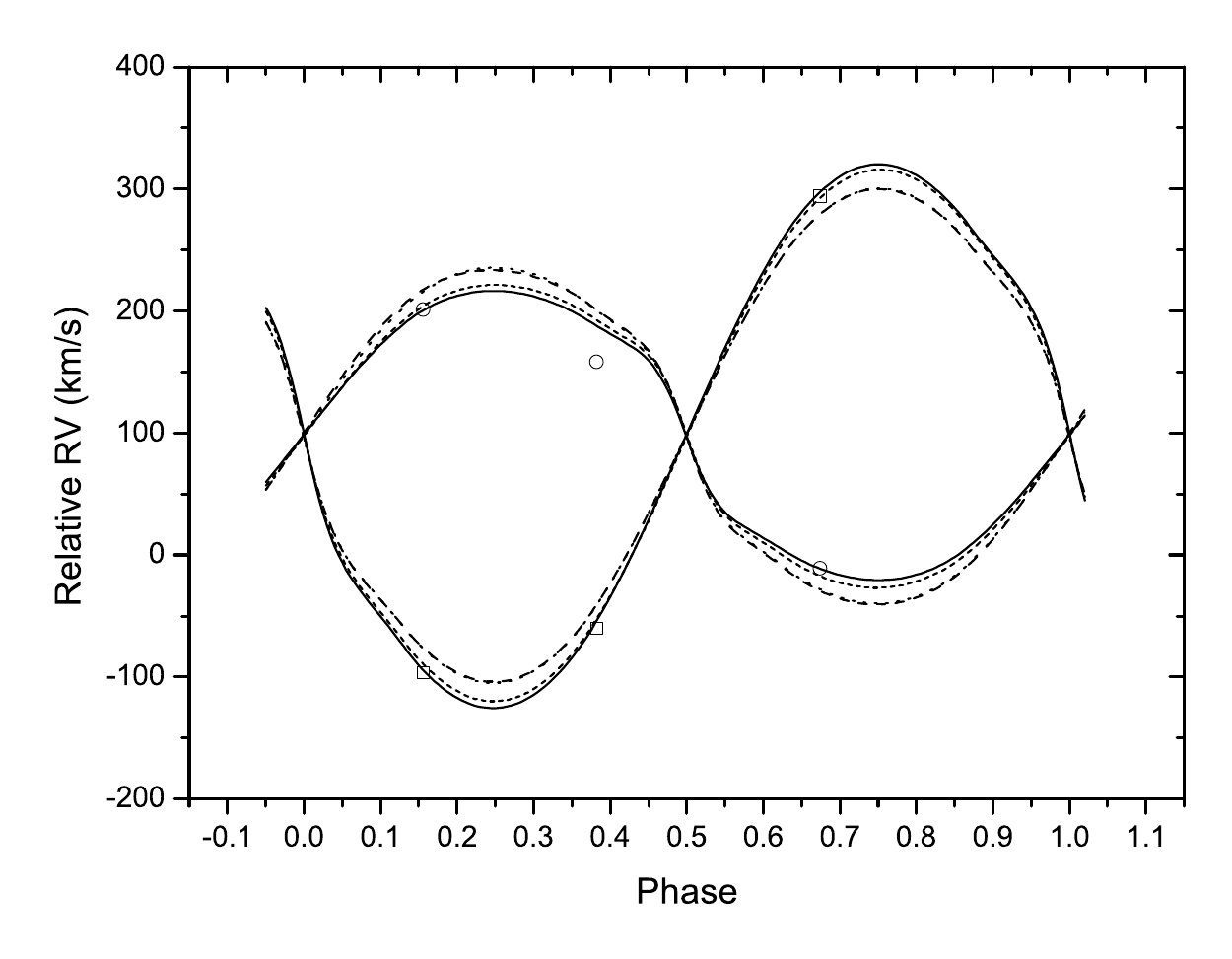}
\caption{
Comparison of the RV data from gaussian fit and RV curves from calculation. Open circles and  squares denote the RV data from gaussian fit, while solid, short-dashed, dashed and dotted lines represent curves from LCs 2016, lc05b, lc18 and lc05a, respectively}
\label{figrv}
\end{center}
\end{figure}

To compare the spectroscopic result with photometric solutions, the W-D program was utilized to calculate the RVs which takes consideration of eclipse-proximity correction (see the manual of W-D program). RV curves from four set of solutions were compared and shown in Figure \ref{figrv}, including the solutions from lc05a (hot spots), lc05b, lc18a based on the K2 data, and the solutions from LCs 2016 (with spots) based on the groud-based telescope. Seen from this figure, both the RV curves from LCs 2016 group and lc05b group can fit the observation well. The main difference between the two sets of solutions is the mass ratio, of which the averaged value is  around 1.76. Then the absolute parameters (masses) were calculated and listed in Table \ref{tababs} (the third row).
 
\begin{figure}
\begin{center}
\includegraphics[angle=0,scale=0.6]{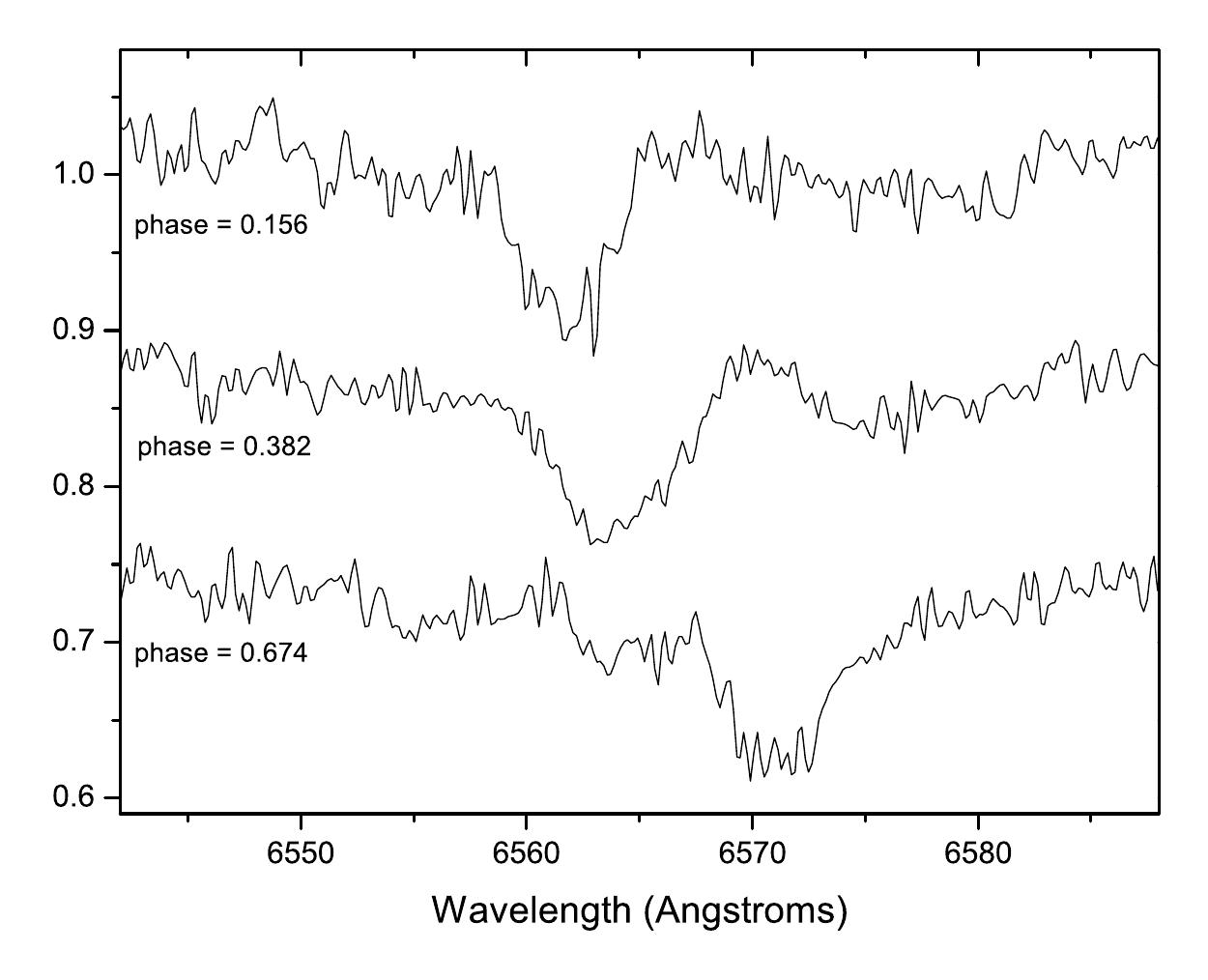}
\caption{H$\alpha$ absorption of ILCnc from msp data. From top to bottom are msp091, msp119 and msp151 respectively.}
\label{figha}
\end{center}
\end{figure}

Here we also take a glance at the H$\alpha$ line of our target from the spectra. Figure \ref{figha} shows how the broad H$\alpha$ absorption changes with time. The absorption line from the primary component (more massive) seems to be missing or even has emission feature (msp119). This is peculiar and might indicate the strong activity on the surface of the binary.

\section{Discussions and Conclusions}
Comprehensive photometric solutions were derived independently both for ground-based data and space-based data. For the light curves from ground-based telescopes, a few cases including spots, a third light and even spot evolution were taken into consideration. For the light curves from K2 survey, cases with spots were mostly considered. The results from all cases indicate that the variable is a W-subtype shallow contact binary with a moderate high orbital inclination ($i = 73.5^{\circ}\pm 0.5^{\circ}$). However, the solutions vary in some parameters, especially the mass ratio $q$. It is hard to figure out which is the correct one simply from the photometric solutions because even all of them give good fit (see Figure \ref{figlckp}). It is suggested that the determination of mass ratio is affected by the spot settings. The spectroscopic investigation based on LAMOST msp data helps to clarify the result.

\subsection{W-subtype Active Contact Binary System}
Based on the results from both photometric solutions and RV analysis, the mass ratio of IL Cnc is calculated to be $q = M_2/M_1 = 1.76\pm 0.05 (1/q = 0.57\pm 0.02)$. This value happens to be within the range that Alton (2018) declared. Together with the temperature difference about $\Delta T = T_2 - T_1 = -280\pm 20$ K and a fill-out factor approximately $9\%$, it is clarified that IL Cnc is a W-subtype (the temperature of the less massive component is higher than that of the more massive component) shallow contact system. The RV curves give a further confirmation for this.
This seems to support that K-type contact binary stars are more likely to be W-subtype. More examples up-to-date are V1799 Ori \citep{Liun14c}, 1SWASP J064501.21+342154.9 \citep{Liun14b}, NSVS 2706134 \citep{Martigi16} and 07g-3-00820 \citep{Gao17}. 

The asymmetry in the light curves was usually modeled by spots on the components which indicates that the system may posses solar-like activities (e.g. \citet{Qian07}). The continuous change of O'Connell effect in the light curves from K2 survey further tells that this system is highly active. The peculiar feature of H$\alpha$ absorption of the spectral data may possibly be caused by spot activities. Another issue to be mentioned is that the errors listed in almost all the photometric solutions are pretty small, compared to the real uncertainties inferred from solutions of all cases. This suggests that errors from single set of solutions need to be adopted prudently.

\begin{figure}[htbp]
\begin{center}
\includegraphics[angle=0,width=8.1cm]{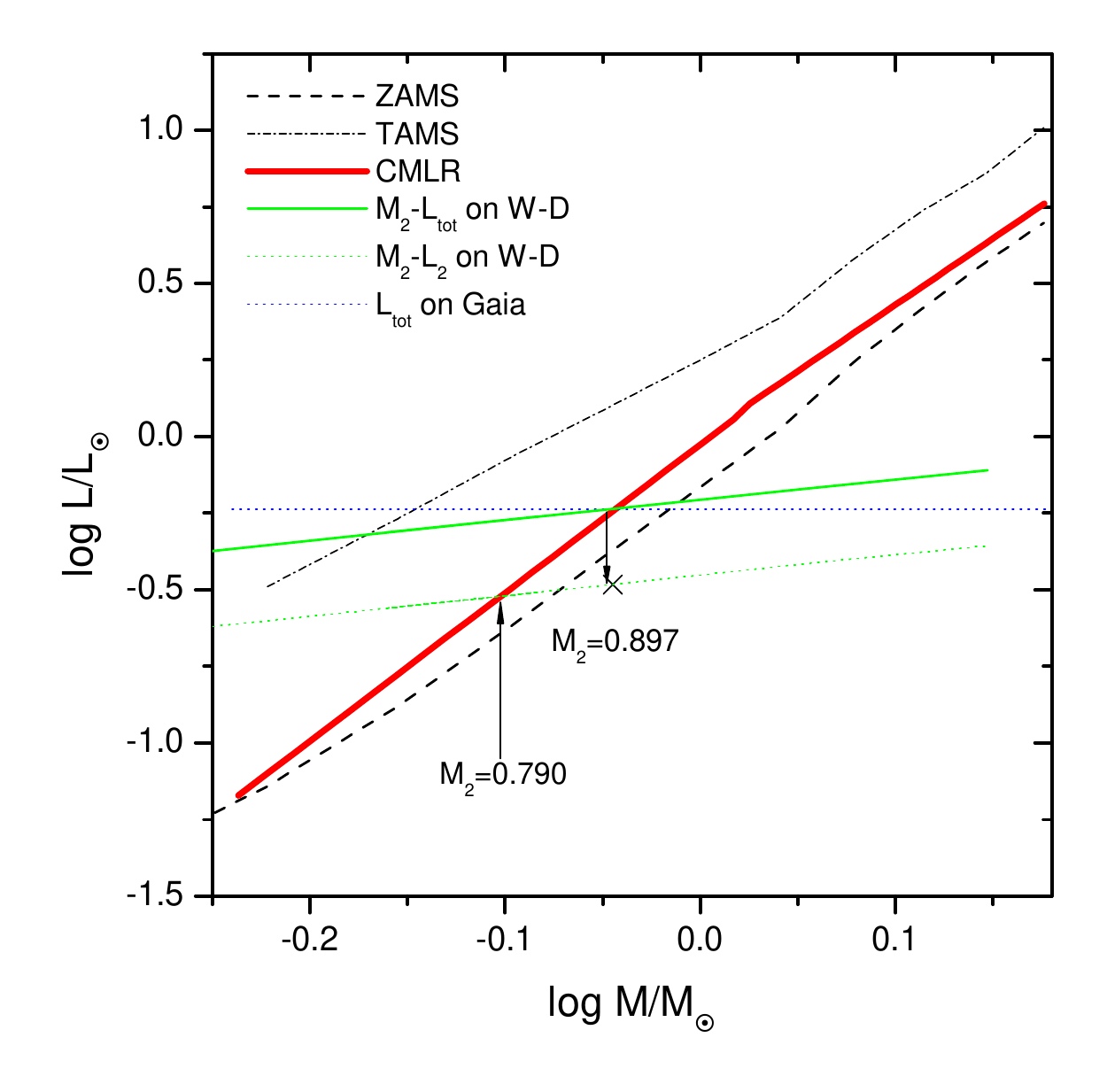}
\caption{
The mass-luminosity diagram of IL Cnc. Dashed and dot dashed lines denote ZAMS and TAMS lines respectively, based on \citet{Girardi00}. The red solid line (thick) denotes a classical mass-luminosity relation (CMLR). The green solid (thin) and dotted lines denote the total luminosity ($L_{tot}$) and the luminosity of star 2 ($L_2$) that computed by W-D code, respectively. The blue dotted line represents the luminosity value based on Gaia parallax data. The cross denote the result from RV data.}
\label{figlum}
\end{center}
\end{figure}

Except for using the RV data as mentioned in Section \ref{secmsp}, the absolute parameters of IL Cnc could also be estimated directly based on the mass-luminosity diagram, which is shown in Figure \ref{figlum}. Taking advantage of the parallax data from Gaia, the luminosity value was calculated with the following formulae (e.g. \citet{Chen18}),
\begin{equation}
M_V = m_V - 5\times\textrm{log}(1000/\pi) + 5 - A_V
\end{equation}
\begin{equation}
M_{bol} = M_V + BC
\end{equation}
\begin{equation}
\textrm{log}{}(L/L_{\solar}) = 0.4\times(4.74-M_{bol})
\end{equation}
where the parallax $\pi = 4.08\pm 0.04$ mas \citep{gaia18}, $m_V$ = 12.6 mag (VSX database\footnote{https://www.aavso.org/vsx/index.php}), the interstellar extinction $A_V \sim 0$ ($b > 30^{\circ}$), and $BC= -0.33$ \citep{Bessell98}. The total luminosity $L_{tot}$ is then estimated to be about 0.579 $\rm L_{\solar}$. By using the W-D code computed mass-luminosity line (based on the spot solutions from LCs 2016, see Table \ref{tabwd}), two values were derived for the mass of IL Cnc based on whether it follows the classical mass-luminosity relation (hereafter CMLR, \citet{Eker15}) or it exactly meets the estimation from Gaia parallax (see Figure \ref{figlum}). The estimated values of absolute parameters are listed in Table \ref{tababs}. Implied from Figure \ref{figlum}, the results from Gaia parallax is in good agreement with that from the RV data. Both of the them show that the primary component of IL Cnc deviates the Mass-Luminosity  relation of zero age main sequence, i.e. they have much lower temperatures. This may suggest that the primary component has significant magnetic activities or there may be strong energy transfer between components. Nevertheless, both hypotheses need more evidence. 

\begin{table}[htpb]
\caption{Estimation of component masses for IL Cnc.}\label{tababs}
\small
\begin{center}
\begin{tabular}{lcc}
\hline
Methods  &  $M_1$ (M$_{\solar}$)   & $M_2$ (M$_{\solar}$)  \\
\hline
on CMLR  &   0.436  &  0.790  \\
on Gaia  &   0.495  &  0.897 \\
on RVs   &   0.512  &  0.902  \\
\hline
\end{tabular}
\end{center}
\end{table}

\subsection{Supplemental comments on the orbital period investigation}\label{seccomper}

The results on orbital period investigation were presented both in Section \ref{secper} and \ref{seckoc}, which are based on only groud-based data and the whole data, respectively. Besides the linear fit, we tried to fit the O-C diagrams with a general equation \citep{Irwin52}:
\begin{equation}\label{eq:epoec}
\begin{array}{ll}%%\begin{split}
\textrm{O-C} = & \Delta T_0 + \Delta P\cdot E+ \frac{\beta}{2} E^2 + \\
                 & A \times [\frac{1-e^2}{1+e\cos\nu}\sin(\nu+w)+ e\sin w]
\end{array}%%\end{split}
\end{equation}
Detailed explanation of the parameters can be found in the paper of \citet{LiaoW10}. A weighted least-squares method was used in the fitting. The results turned out that only the groud-based data could have a rough solution. After the K2 data were employed, no meaningful solution was found, which probably suggests that no eccentric LITE (light-time effect, see \citet{Fries73}) solution works here. Taking account that spot activities might affect the determination of the real conjunction of eclipse, we tried to fit the overall O-C diagram separately for the primary and secondary minima. Finally, the periodic solution was only found for the primary data. According to this solution, the orbital period changes with a period about 10.6 yr. 

One explanation for the cyclic variation might be the light travel time effect (sometimes called "light-time effect"). Using a well known method (an example can be found in \citet{Liun15}) based on the periodic solution, the mass function was calculated to be $f(m) = 8.8(\pm 2.6)\times 10^{-5}$ M$_{\solar}$ and the minimum mass of an additional body was then derived to be $m_3~(i'= 90^{\circ}) = 0.061 \pm 0.006$ M$_{\solar}$. However, the difficulty of this hypothesis here is that the overall O-C diagram has a large scatter which could not be well fitted by the periodic equation and the cyclic variation only matches the primary minima.

Another hypothesis seems to be more convincing, which claims that the oscillation may be caused by the spot activity mechanism. Recently, \citet{Tran13} reported that the O-C curves of short period binary stars display quasi-periodicities with typical amplitudes of 200$\sim$300 s due to spot activities. \citet{Balaji15} even found a new method for tracking the phase of modulations in very short period binary systems, which extracts more information about starspot and confirmed the hypothesis of Tran et al.. The remarkable O'Connell effect shown in Figure \ref{figconn} helps to demonstrate that the spot activity mechanism may be more plausible for the cyclic variation. 
It should be mentioned that on the opposite, \citet{Rappaport13} reported a few candidate triple systems based on ETV (eclipsing time variation) data, a few of which have amplitudes below hundreds of seconds. This suggests that LITE explanation is not rejected yet, taking account of the relatively large errors ($\sim$ 0.001 days) of the early timing data. To confirm the main reason of the cyclic variation here (tentatively spot activity mechanism), more precise data (especially continuous data) might be required in the future.

For secular variation of the orbital period, the results from the O$-$C diagrams suggest very small values (the period change rate (absolute value) is determined to be only $7(\pm 5) \times 10^{-9}$ days $\cdot$ yr$^{-1}$ from Eq. \ref{eq:epocy}). This is quite similar to some early K-type systems such as V1799 Ori and RV CVn \citep{Liun14a}, which indicates that this phenomenon may be prevalent among K-type contact binaries. 

\begin{ack}
This work is supported by the Chinese National Natural Science Foundation (Grant No. 11503077), and partly supported by the research fund of NARIT and the West Light Foundation of Chinese Academy of Sciences. New CCD photometric data were obtained with the 2.4-m telescope of Thai National Observatory, the sino-Thai 70 cm telescope in Lijiang and 60 cm telescope of Yunnan Observatories. The K2 data was downloaded from MAST database. The spectral data were retrieved from LAMOST DR6 and DR7 database. This work also makes use of data from Simbad, VSX, Vizier, and Gaia databases. This work is part of the research activities at the National Astronomical Research Institute of Thailand (Public Organization). We would also thank Prof. Qian in Yunnan Observatories and Prof. Soonthornthum in National Astronomical Research Institute of Thailand for valuable suggestions and help. Anonymous colleagues aslo provide us some important technique support. We are grateful to the anonymous referee for valuable advices which have improved the manuscript greatly. 
\end{ack}

\appendix
\setcounter{table}{0}
\renewcommand{\thetable}{A\arabic{table}}

\section*{Newly determined minimum light times from Kepler data}
\begin{longtable}{ccccccccc}
  \caption{Newly determined minimum light times from Kepler data. NA: number of data for determination}\label{tabkpmin}
  \hline              
  BJD & Err & NA &  BJD & Err & NA &   BJD & Err & NA   \\ 
  2,400,000+  & (days) &  &  2,400,000+  & (days) &  &  2,400,000+  & (days) &  \\
\endfirsthead
  \hline
  BJD & Err & NA &  BJD & Err & NA &   BJD & Err & NA   \\ 
  2,400,000+  & (days) &  &  2,400,000+  & (days) &  &  2,400,000+  & (days) &  \\
  \hline
\endhead
  \hline
\endfoot
  \hline
\endlastfoot
  \hline
   57140.56957 &  0.00019 &  15  &  57181.78854 &  0.00017 &  19  &  58261.51256 &  0.00021 &  21 \\
   57140.43590 &  0.00014 &  15  &  57181.65495 &  0.00020 &  19  &  58261.37840 &  0.00015 &  21 \\
   57140.83712 &  0.00017 &  19  &  57182.59153 &  0.00017 &  20  &  58262.31540 &  0.00021 &  17 \\
   57140.70340 &  0.00010 &  15  &  57182.45783 &  0.00019 &  18  &  58262.44884 &  0.00017 &  18 \\
   57141.64014 &  0.00016 &  21  &  57183.39450 &  0.00017 &  20  &  58263.11842 &  0.00019 &  21 \\
   57141.77382 &  0.00017 &  20  &  57183.52878 &  0.00011 &  15  &  58263.25202 &  0.00018 &  20 \\
   57142.44314 &  0.00018 &  21  &  57184.19745 &  0.00017 &  20  &  58263.65400 &  0.00019 &  17 \\
   57142.57676 &  0.00016 &  20  &  57184.06382 &  0.00022 &  19  &  58263.78726 &  0.00016 &  17 \\
   57143.24614 &  0.00018 &  21  &  57185.00034 &  0.00016 &  16  &  58264.72467 &  0.00019 &  17 \\
   57143.11228 &  0.00014 &  15  &  57184.86685 &  0.00021 &  19  &  58264.59023 &  0.00013 &  15 \\
   57144.04914 &  0.00018 &  21  &  57185.80341 &  0.00021 &  20  &  58265.52753 &  0.00016 &  17 \\
   57144.18266 &  0.00018 &  20  &  57185.93752 &  0.00023 &  19  &  58265.66097 &  0.00015 &  18 \\
   57144.85219 &  0.00012 &  16  &  57186.60637 &  0.00020 &  20  &  58266.33033 &  0.00017 &  16 \\
   57144.98560 &  0.00018 &  20  &  57186.74065 &  0.00021 &  20  &  58266.46393 &  0.00014 &  20 \\
   57145.65536 &  0.00014 &  19  &  57187.40934 &  0.00019 &  20  &  58267.13329 &  0.00020 &  20 \\
   57145.78856 &  0.00018 &  20  &  57187.27600 &  0.00025 &  19  &  58267.26690 &  0.00015 &  20 \\
   57146.45832 &  0.00014 &  20  &  57188.21234 &  0.00016 &  21  &  58267.93622 &  0.00017 &  20 \\
   57146.59145 &  0.00016 &  20  &  57188.07900 &  0.00028 &  19  &  58267.80206 &  0.00016 &  18 \\
   57147.26128 &  0.00011 &  15  &  57189.01539 &  0.00018 &  19  &  58268.73925 &  0.00012 &  15 \\
   57147.39440 &  0.00016 &  20  &  57189.14997 &  0.00020 &  15  &  58268.60500 &  0.00015 &  19 \\
   57148.06407 &  0.00018 &  20  &  57189.81841 &  0.00020 &  19  &  58269.54204 &  0.00016 &  19 \\
   57148.19736 &  0.00017 &  20  &  57189.68500 &  0.00019 &  20  &  58269.67562 &  0.00017 &  19 \\
   57148.86698 &  0.00018 &  19  &  57190.88907 &  0.00017 &  19  &  58270.34503 &  0.00016 &  18 \\
   57148.73286 &  0.00010 &  15  &  57190.48784 &  0.00021 &  16  &  58270.47871 &  0.00016 &  20 \\
   57149.66993 &  0.00017 &  19  &  57191.42437 &  0.00016 &  20  &  58271.14808 &  0.00017 &  17 \\
   57149.53571 &  0.00017 &  20  &  57191.29098 &  0.00022 &  19  &  58271.01405 &  0.00016 &  20 \\
   57150.47279 &  0.00017 &  18  &  57192.22741 &  0.00019 &  19  &  58271.95113 &  0.00019 &  21 \\
   57150.33850 &  0.00017 &  15  &  57192.36160 &  0.00020 &  19  &  58271.81707 &  0.00015 &  20 \\
   57151.27579 &  0.00017 &  19  &  57193.03054 &  0.00018 &  18  &  58272.75421 &  0.00019 &  20 \\
   57151.14202 &  0.00013 &  15  &  57192.89686 &  0.00019 &  19  &  58272.62005 &  0.00014 &  15 \\
   57152.07878 &  0.00020 &  19  &  57194.10102 &  0.00020 &  19  &  58273.55719 &  0.00021 &  20 \\
   57151.94461 &  0.00018 &  20  &  57193.69978 &  0.00019 &  18  &  58273.42316 &  0.00018 &  17 \\
   57152.88184 &  0.00014 &  16  &  57194.63608 &  0.00019 &  16  &  58274.36020 &  0.00021 &  16 \\
   57152.74750 &  0.00017 &  19  &  57194.77038 &  0.00017 &  18  &  58274.22612 &  0.00024 &  19 \\
   57153.68493 &  0.00018 &  20  &  57195.43944 &  0.00017 &  17  &  58275.16311 &  0.00021 &  20 \\
   57153.55053 &  0.00018 &  20  &  57195.57331 &  0.00017 &  19  &  58275.02911 &  0.00025 &  19 \\
   57154.48791 &  0.00019 &  20  &  57196.24225 &  0.00018 &  19  &  58275.96591 &  0.00018 &  17 \\
   57154.62121 &  0.00017 &  20  &  57196.10866 &  0.00020 &  18  &  58276.09975 &  0.00018 &  17 \\
   57155.29102 &  0.00012 &  15  &  57197.04510 &  0.00017 &  15  &  58276.76926 &  0.00013 &  17 \\
   57155.15650 &  0.00018 &  20  &  57196.91201 &  0.00013 &  17  &  58276.63499 &  0.00017 &  20 \\
   57156.09388 &  0.00017 &  21  &  57197.84812 &  0.00016 &  19  &  58277.57197 &  0.00022 &  20 \\
   57155.95948 &  0.00019 &  20  &  57197.98240 &  0.00019 &  20  &  58277.43794 &  0.00016 &  20 \\
   57156.89686 &  0.00018 &  20  &  57198.65108 &  0.00018 &  19  &  58278.37491 &  0.00020 &  20 \\
   57156.76261 &  0.00011 &  15  &  57198.51769 &  0.00018 &  20  &  58278.50855 &  0.00016 &  20 \\
   57157.69981 &  0.00017 &  20  &  57199.45402 &  0.00019 &  19  &  58279.17797 &  0.00022 &  21 \\
   57157.83314 &  0.00017 &  21  &  57199.32051 &  0.00017 &  18  &  58279.31144 &  0.00015 &  21 \\
   57158.50277 &  0.00019 &  20  &  57199.98933 &  0.00018 &  19  &  58279.98096 &  0.00020 &  21 \\
   57158.36853 &  0.00018 &  19  &  57200.39120 &  0.00017 &  18  &  58279.84683 &  0.00019 &  19 \\
   57159.30578 &  0.00019 &  20  &  57201.06002 &  0.00014 &  16  &  58280.78408 &  0.00020 &  21 \\
   57159.17153 &  0.00019 &  19  &  57201.19417 &  0.00017 &  18  &  58280.64990 &  0.00016 &  15 \\
   57160.10872 &  0.00018 &  20  &  57201.86311 &  0.00017 &  18  &  58281.58704 &  0.00020 &  21 \\
   57160.24217 &  0.00017 &  19  &  57201.72958 &  0.00019 &  19  &  58281.72059 &  0.00015 &  15 \\
   57160.91169 &  0.00015 &  15  &  57202.66606 &  0.00017 &  19  &  58282.38999 &  0.00017 &  18 \\
   57160.77750 &  0.00016 &  20  &  57202.80022 &  0.00017 &  19  &  58282.25578 &  0.00013 &  21 \\
   57161.71465 &  0.00018 &  20  &  57203.46931 &  0.00015 &  15  &  58283.19297 &  0.00020 &  21 \\
   57161.58047 &  0.00017 &  20  &  57203.60322 &  0.00016 &  19  &  58283.05876 &  0.00013 &  21 \\
   57162.51758 &  0.00017 &  21  &  57204.27205 &  0.00017 &  20  &  58283.99583 &  0.00021 &  20 \\
   57162.65119 &  0.00015 &  16  &  57204.40607 &  0.00016 &  15  &  58284.12939 &  0.00015 &  20 \\
   57163.32064 &  0.00013 &  15  &  57205.07503 &  0.00017 &  20  &  58284.79911 &  0.00020 &  15 \\
   57163.45404 &  0.00017 &  21  &  57204.94182 &  0.00009 &  15  &  58284.93229 &  0.00016 &  21 \\
   57164.12345 &  0.00016 &  21  &  57205.87795 &  0.00016 &  21  &  58285.60180 &  0.00018 &  20 \\
   57163.98918 &  0.00017 &  16  &  57206.01233 &  0.00016 &  19  &  58285.46756 &  0.00019 &  20 \\
   57164.92639 &  0.00016 &  21  &  57206.68091 &  0.00018 &  20  &  58286.40481 &  0.00017 &  20 \\
   57164.79262 &  0.00014 &  15  &  57206.81519 &  0.00017 &  16  &  58286.53818 &  0.00017 &  19 \\
   57165.72934 &  0.00015 &  21  &  57207.48391 &  0.00017 &  20  &  58287.20777 &  0.00017 &  19 \\
   57165.86299 &  0.00016 &  18  &  57207.35093 &  0.00010 &  15  &  58287.34110 &  0.00017 &  20 \\
   57166.53230 &  0.00017 &  21  &  57208.28686 &  0.00014 &  20  &  58288.01076 &  0.00018 &  20 \\
   57166.39832 &  0.00018 &  20  &  57208.15368 &  0.00019 &  19  &  58288.14405 &  0.00017 &  20 \\
   57167.33526 &  0.00018 &  21  &  57209.08990 &  0.00011 &  16  &  58288.81370 &  0.00018 &  20 \\
   57167.20129 &  0.00019 &  19  &  57209.22432 &  0.00018 &  19  &  58288.67941 &  0.00014 &  15 \\
   57168.13818 &  0.00018 &  21  &  57209.89274 &  0.00016 &  20  &  58289.61654 &  0.00021 &  15 \\
   57168.27197 &  0.00019 &  19  &  57210.02728 &  0.00017 &  19  &  58289.74997 &  0.00016 &  20 \\
   57168.94101 &  0.00013 &  15  &  57210.69573 &  0.00017 &  20  &  58290.41984 &  0.00019 &  16 \\
   57169.07490 &  0.00015 &  19  &  57210.56264 &  0.00015 &  19  &  58290.55292 &  0.00018 &  20 \\
   57169.74408 &  0.00017 &  21  &  57211.49884 &  0.00011 &  15  &  58291.22271 &  0.00022 &  21 \\
   57169.61019 &  0.00015 &  17  &  57211.36564 &  0.00014 &  19  &  58291.08830 &  0.00018 &  21 \\
   57170.54707 &  0.00017 &  21  &  57212.30167 &  0.00017 &  20  &  58292.02546 &  0.00018 &  16 \\
   57170.68089 &  0.00017 &  18  &  57212.43630 &  0.00014 &  19  &  58292.15894 &  0.00015 &  20 \\
   57171.35003 &  0.00018 &  20  &  57213.10464 &  0.00017 &  20  &  58292.82884 &  0.00015 &  17 \\
   57171.48390 &  0.00017 &  18  &  57212.97165 &  0.00014 &  16  &  58292.96191 &  0.00015 &  20 \\
   57172.15300 &  0.00018 &  20  &  58252.41230 &  0.00014 &  16  &  58293.63159 &  0.00020 &  21 \\
   57172.28690 &  0.00018 &  19  &  58252.54602 &  0.00023 &  15  &  58293.76491 &  0.00016 &  20 \\
   57172.95597 &  0.00018 &  20  &  58252.67984 &  0.00013 &  17  &  58294.43453 &  0.00020 &  21 \\
   57172.82248 &  0.00011 &  15  &  58252.81353 &  0.00019 &  17  &  58294.56778 &  0.00013 &  18 \\
   57173.75895 &  0.00017 &  21  &  58253.48268 &  0.00018 &  19  &  58295.23748 &  0.00019 &  21 \\
   57173.89294 &  0.00016 &  21  &  58253.34883 &  0.00016 &  20  &  58295.37087 &  0.00014 &  20 \\
   57174.56190 &  0.00016 &  21  &  58254.28568 &  0.00019 &  19  &  58296.04054 &  0.00014 &  20 \\
   57174.42826 &  0.00018 &  21  &  58254.41948 &  0.00014 &  19  &  58296.17384 &  0.00014 &  20 \\
   57175.36486 &  0.00016 &  21  &  58255.08872 &  0.00020 &  20  &  58296.84348 &  0.00014 &  20 \\
   57175.23157 &  0.00015 &  15  &  58255.22238 &  0.00012 &  20  &  58296.70928 &  0.00012 &  15 \\
   57176.16785 &  0.00017 &  21  &  58255.89168 &  0.00020 &  20  &  58297.64642 &  0.00013 &  20 \\
   57176.30187 &  0.00017 &  20  &  58255.75755 &  0.00011 &  16  &  58297.77960 &  0.00022 &  20 \\
   57176.97097 &  0.00013 &  15  &  58256.69457 &  0.00019 &  17  &  58298.44923 &  0.00015 &  18 \\
   57177.10481 &  0.00016 &  19  &  58256.56077 &  0.00009 &  15  &  58298.58259 &  0.00022 &  20 \\
   57177.77378 &  0.00017 &  20  &  58257.49769 &  0.00018 &  21  &  58299.25221 &  0.00019 &  20 \\
   57177.90782 &  0.00016 &  20  &  58257.63130 &  0.00014 &  20  &  58299.11824 &  0.00019 &  15 \\
   57178.57673 &  0.00015 &  20  &  58258.30071 &  0.00017 &  16  &  58300.05514 &  0.00018 &  20 \\
   57178.44320 &  0.00016 &  19  &  58258.43421 &  0.00015 &  19  &  58299.92087 &  0.00017 &  19 \\
   57179.37973 &  0.00013 &  15  &  58259.10364 &  0.00019 &  21  &  58300.85823 &  0.00012 &  15 \\
   57179.24622 &  0.00016 &  19  &  58258.96957 &  0.00014 &  21  &  58300.72381 &  0.00017 &  18 \\
   57180.18260 &  0.00017 &  20  &  58259.90641 &  0.00018 &  16  &  58301.39340 &  0.00018 &  15 \\
   57180.31684 &  0.00015 &  19  &  58260.04015 &  0.00016 &  21  &  58301.52670 &  0.00016 &  15 \\
   57180.98558 &  0.00017 &  20  &  58260.70987 &  0.00016 &  17  &  &  &  \\
   57180.85216 &  0.00013 &  16  &  58260.84312 &  0.00016 &  21  &  &  &  \\
\end{longtable}

%%%
% See the manual for the detail.
%%%

\end{document}